\documentclass[superscriptaddress,twocolumn,showpacs,
amssymb,amsmath,longbiliography,nobibnotes,aps,prd,
nofootinbib]{revtex4-2}
\pdfoutput=1
\usepackage{graphicx,subfigure,bm,color,psfrag,hyperref}
\usepackage{amsfonts}
\usepackage{lipsum}
\usepackage{mathtools}
\usepackage{verbatim}
\usepackage[normalem]{ulem}
\usepackage[dvipsnames]{xcolor}
\hypersetup{colorlinks,linkcolor={blue},citecolor={red},urlcolor={greenW}} 
\definecolor{greenW}{rgb}{0.0, 0.55, 0.1}

\begin{document}

\title{Dark Energy Is Not That Into You: Variable Couplings after DESI DR2 BAO}

\author{Weiqiang Yang}
\email{d11102004@163.com}
\affiliation{Department of Physics, Liaoning Normal University, Dalian, 116029, People's Republic of China}

\author{Sibo Zhang}
\email{sbzhang02@163.com}
\affiliation{Department of Physics, Liaoning Normal University, Dalian, 116029, People's Republic of China}

\author{Olga Mena}
\email{omena@ific.uv.es}
\affiliation{Instituto de F\'{i}sica Corpuscular (CSIC-Universitat de Val\`{e}ncia), E-46980 Paterna, Spain} 

\author{Supriya Pan}
\email{supriya.maths@presiuniv.ac.in}
\affiliation{Department of Mathematics, Presidency University, 86/1 College Street, Kolkata 700073, India}
\affiliation{Department of Mathematics, Faculty of Applied Sciences, Durban University of Technology, Durban 4000, Republic of South Africa}

\author{Eleonora Di Valentino}
\email{e.divalentino@sheffield.ac.uk}
\affiliation{School of Mathematical and Physical Sciences, University of Sheffield, Hounsfield Road, Sheffield S3 7RH, United Kingdom}

\begin{abstract}
In interacting dark energy (DE) and dark matter (DM) scenarios, the interaction function typically includes a coupling parameter $\xi$ that quantifies the strength of energy exchange between the dark sectors. While $\xi$ is often assumed to be constant, there is no fundamental reason to exclude a time-dependent coupling, which could provide a more general and realistic description of dark sector dynamics. In this work, we study two widely used interacting models involving pressureless DM and DE, where the coupling parameter is allowed to vary with the scale factor $a$. Specifically, we consider two parametrizations: $\xi(a) = \xi_0 + \xi_a (1-a)$ and $\xi(a) = \xi_0 \left(1  + \frac{1-a}{a^2 + (1-a)^2} \right)$, and constrain them using the latest cosmological observations, including Planck 2018 CMB data, DESI DR2 BAO measurements, and multiple Type Ia supernovae samples. Our results show that one scenario yields evidence for a non-zero interaction at more than 95\% confidence level, while the remaining cases indicate at most mild or inconclusive signs of interaction. These findings highlight the potential of variable coupling models and the importance of continued investigation into the nature of the dark sectors.
\end{abstract}

\keywords{Dark matter and dark energy, interacting cosmologies, cosmological observations}
\maketitle

\section{Introduction}
\label{sec-intro}

The concordance cosmological paradigm, $\Lambda$-Cold Dark Matter (hereafter $\Lambda$CDM), built within the framework of Einstein's General Relativity (GR), has been quite successful in explaining a large number of astronomical datasets. In this well-known setup, $\Lambda$ corresponds to the vacuum energy with a constant equation of state (EoS) of $-1$, and DM, by default, is pressureless. Furthermore, CDM and $\Lambda$ in this framework remain independently conserved—an assumption that is theoretically motivated. 
This theoretical assumption can be relaxed by introducing an interaction between these two components; that is, by allowing an energy-momentum exchange mechanism between the dark sectors. This framework is known as interacting dark energy (DE), interacting dark energy–dark matter (DE–DM), or more broadly, as interacting cosmologies. These interacting cosmologies were originally motivated by the limitations of the $\Lambda$CDM cosmological model, and shortly after their introduction, this special class of cosmological theory became very popular in the cosmology community~\cite{Amendola:1999er,Amendola:2003eq,Cai:2004dk,Huey:2004qv,Gumjudpai:2005ry,Brookfield:2005td,Berger:2006db,delCampo:2006vv,Barrow:2006hia,delCampo:2008sr,Valiviita:2008iv,Chongchitnan:2008ry,Gavela:2009cy,Xia:2009zzb,Gavela:2010tm,Mena:2010zz,Baldi:2010td,LopezHonorez:2010esq,Lee:2011tq,Beynon:2011hw,Amendola:2011ie,Farajollahi:2012zz,Pettorino:2012ts,Chimento:2012aea,Salvatelli:2013wra,Pettorino:2013oxa,Xia:2013nua,Li:2013bya,Yang:2014okp,Yang:2014gza,Salvatelli:2014zta,Wang:2014xca,Li:2014eba,Pan:2012ki,Li:2015vla,Cui:2015ueu,Landim:2015uda,Yang:2016evp,Wang:2016lxa,Nunes:2016dlj,vandeBruck:2016hpz,Xia:2016vnp,Pan:2016ngu,Fay:2016yow,Kumar:2017dnp,DiValentino:2017iww,Sharov:2017iue,Yang:2017yme,Yang:2017ccc,Pan:2017ent,Mifsud:2017fsy,VanDeBruck:2017mua,Yang:2017zjs,Yang:2018euj,Yang:2018xlt,Li:2018jiu,Barros:2019rdv,Teixeira:2019tfi,DiValentino:2019jae,Paliathanasis:2019hbi,Yang:2019uog,Pan:2019gop,Nakamura:2019phn,DiValentino:2019ffd,Yang:2019bpr,Kumar:2019wfs,Martinelli:2019dau,Costa:2019uvk,Arevalo:2019axj,Pan:2019jqh,Cheng:2019bkh,Oikonomou:2019nmm,Kase:2019veo,Pan:2020zza,Gomez-Valent:2020mqn,DiValentino:2020leo,Yang:2020uga,Lucca:2020zjb,DiValentino:2020kpf,Hogg:2020rdp,Yao:2020hkw,BeltranJimenez:2020qdu,Yao:2020pji,Lucca:2021dxo,Lucca:2021eqy,Nunes:2021zzi,Gariazzo:2021qtg,Potting:2021bje,Sa:2021eft,daFonseca:2021imp,Gao:2021xnk,Thipaksorn:2022yul,Gomez-Valent:2022bku,Harko:2022unn,Yengejeh:2022tpa,Nunes:2022bhn,Landim:2022jgr,Yao:2022kub,Li:2023gtu,Giare:2024ytc,Kritpetch:2024rgi,Giare:2024smz,Li:2024qso,Halder:2024gag,Carrion:2024jur,Giani:2024nnv,Tsedrik:2025cwc,Liu:2025pxy,Zhai:2025hfi,Li:2025owk,Silva:2025hxw,You:2025uon,Aoki:2025bmj,Yang:2025ume,Abedin:2025yru,Pan:2025qwy,Liu:2025vda,Yang:2025boq,Li:2025ula,Yan:2025iga,Wang:2025znm}. 

In interacting DE–DM theories, a pivotal role is played by the interaction function, which characterizes the rate of energy transfer between the dark sectors. This coupling function involves a dimensionless coupling parameter, $\xi$, which controls the strength of the interaction. Usually, the coupling parameter $\xi$ is assumed to be independent of cosmic evolution, although there is no underlying theory that excludes the possibility of a time-dependent coupling. Indeed, a time-dependent coupling parameter may offer a more general description of the interacting dynamics. 
As the sensitivity of astronomical datasets continues to increase, it is logical to let observations determine whether the coupling parameter—if favored by the data—is dynamical throughout the cosmic expansion history. In the present work, we explore the possibility of a time-dependent coupling, constraining the resulting scenarios using the latest observational datasets from diverse astronomical surveys. 

The article is structured as follows. In Section~\ref{sec-efe}, we present the basic gravitational equations and the models we aim to study. In Section~\ref{sec-data}, we describe the astronomical datasets used to constrain the interacting scenarios. Section~\ref{sec-results} reports the results and their implications. Finally, Section~\ref{sec-summary} provides the overall summary and conclusions.

\section{Interacting dark sectors}
\label{sec-efe}

We shall consider the following scenario: the gravitational sector of the universe is described by GR, the matter sector is minimally coupled to GR, and the geometry of the universe on large scales is described by the spatially flat Friedmann–Lema\^{i}tre–Robertson–Walker (FLRW) line element given by
\begin{equation}
ds^2 = -dt^2 + a^2(t) (dx^2 + dy^2 + dz^2),
\end{equation}
where $a(t)$ is the scale factor of the universe as a function of cosmic time $t$, $(x, y, z)$ are the spatial coordinates, and $(t, x, y, z)$ are the comoving coordinates.
In the matter sector, we consider that DM and DE are not independently conserved; that is, an energy exchange mechanism is allowed between them. Here, we consider DM to be pressureless and DE to correspond to the cosmological constant.
From the total conservation equation of DM+DE, namely,
\begin{equation}
\nabla^\mu (T^{\rm CDM}_{\mu\nu} + T^{\rm DE}_{\mu\nu}) = 0,
\end{equation}
one can explicitly write down the separate conservation equations of the dark sectors as
\begin{eqnarray}
&&\dot{\rho}_x = Q, \label{cons1}\\
&&\dot{\rho}_c = - 3 H \rho_c - Q, \label{cons-dm-de}
\end{eqnarray}
where $\rho_x$ and $\rho_c$ are respectively the energy densities of DE and CDM, the dot denotes the derivative with respect to cosmic time, 
$H$ is the Hubble rate of the FLRW universe, and $Q$ is the rate of energy flow between the dark components. The Hubble rate includes the sum of all energy densities present in the universe and satisfies
\begin{equation}
H^2 = \kappa^2 \sum_{i} \rho_i,
\end{equation}
where $\kappa^2$ is Einstein's gravitational constant and $\rho_i$ is the energy density of the $i$-th fluid present in the energy-momentum tensor of the matter distribution.
In this article, we consider the following two interaction functions~\cite{Yang:2019uzo} 
\begin{eqnarray}
&\rm{IVS1}:& Q = 3 \xi(a) H \rho_{x}, \label{ivs1}\\
&\rm{IVS2}:& Q = 3 \xi(a) H \frac{\rho_c \rho_{x}}{\rho_c + \rho_{x}}, \label{ivs2}
\end{eqnarray}
where $\xi(a)$ is a time-dependent dimensionless coupling. The choice of $\xi(a)$ is not unique, and therefore, in principle, one can construct an infinite number of possibilities. One natural choice for $\xi(a)$ is motivated by the Taylor expansion of $\xi(a)$ around the present time ($a = a_0 = 1$), leading to~\cite{Yang:2019uzo}
\begin{align}\label{Taylor}
\xi(a) = \xi_0 + (a - 1) \xi^{\prime}(a_0 = 1) + \frac{(a - 1)^2}{2!} \xi^{\prime\prime}(a_0 = 1) + \dots
\end{align}
where the prime denotes a derivative with respect to the scale factor. Neglecting the higher-order terms in the above expression, a possible choice for the coupling function can be made as follows~\cite{Yang:2019uzo}:
\begin{eqnarray}\label{xi(a)-1}
\xi(a) = \xi_0 + \xi_a \; (1 - a)~,
\end{eqnarray}
where $\xi_0$ and $\xi_a$ are free parameters. Here, $\xi_0$ refers to the present-day value of $\xi(a)$, while $\xi_a$ characterizes the dynamical nature of the coupling parameter. 
Along with the above choice, we are also interested in coupling functions that are dynamical but involve only one free parameter.
A general construction of such a coupling function follows
\begin{eqnarray}
\xi(a) = \xi_0 f(a),
\end{eqnarray}
where, for $a \rightarrow \infty$ and $a \rightarrow 0$, $f(a)$ should remain finite. Otherwise, the resulting interaction function does not lead to a physically realistic scenario, since it should be well defined throughout the entire evolution of the universe. There are several choices one can make, and here we propose the following:
\begin{eqnarray}\label{xi(a)-2}
\xi(a) = \xi_0 \left(1 + \frac{1 - a}{a^2 + (1 - a)^2}\right),
\end{eqnarray}
where $\xi_0$ is the only free parameter, and $\xi(a)$ is divergence-free for both $a \rightarrow 0$ and $a \rightarrow \infty$. This form is motivated by the Barboza–Alcaniz parametrization in DE~\cite{Barboza:2008rh}. 

Therefore, given the choices for $\xi(a)$ in Eqs.~(\ref{xi(a)-1}) and~(\ref{xi(a)-2}), one can explore the dynamics of the interacting scenarios. We label the scenarios as follows: 
\begin{itemize}
    \item[{\bf \texttt{(i)}}] Interaction function $Q$ in Eq.~(\ref{ivs1}) with $\xi(a)$ from Eq.~(\ref{xi(a)-1}) is labeled as {\bf IVS1a};
    \item[{\bf \texttt{(ii)}}] Interaction function $Q$ in Eq.~(\ref{ivs1}) with $\xi(a)$ from Eq.~(\ref{xi(a)-2}) is labeled as {\bf IVS1b};
    \item[{\bf \texttt{(iii)}}] Interaction function $Q$ in Eq.~(\ref{ivs2}) with $\xi(a)$ from Eq.~(\ref{xi(a)-1}) is labeled as {\bf IVS2a};
    \item[{\bf \texttt{(iv)}}] Interaction function $Q$ in Eq.~(\ref{ivs2}) with $\xi(a)$ from Eq.~(\ref{xi(a)-2}) is labeled as {\bf IVS2b}.
\end{itemize}

\begin{figure*}
    \centering
    \includegraphics[width=0.45\textwidth]{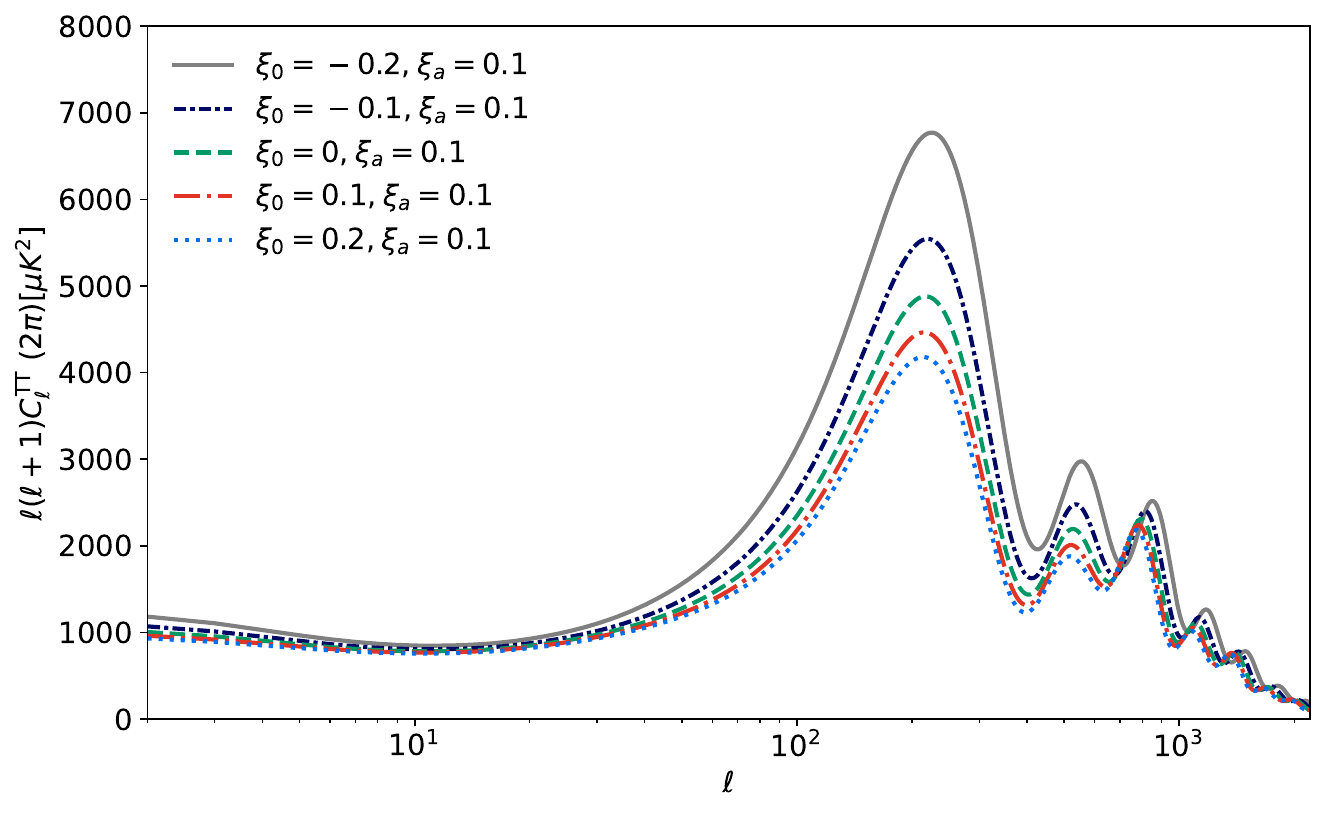}
    \includegraphics[width=0.45\textwidth]{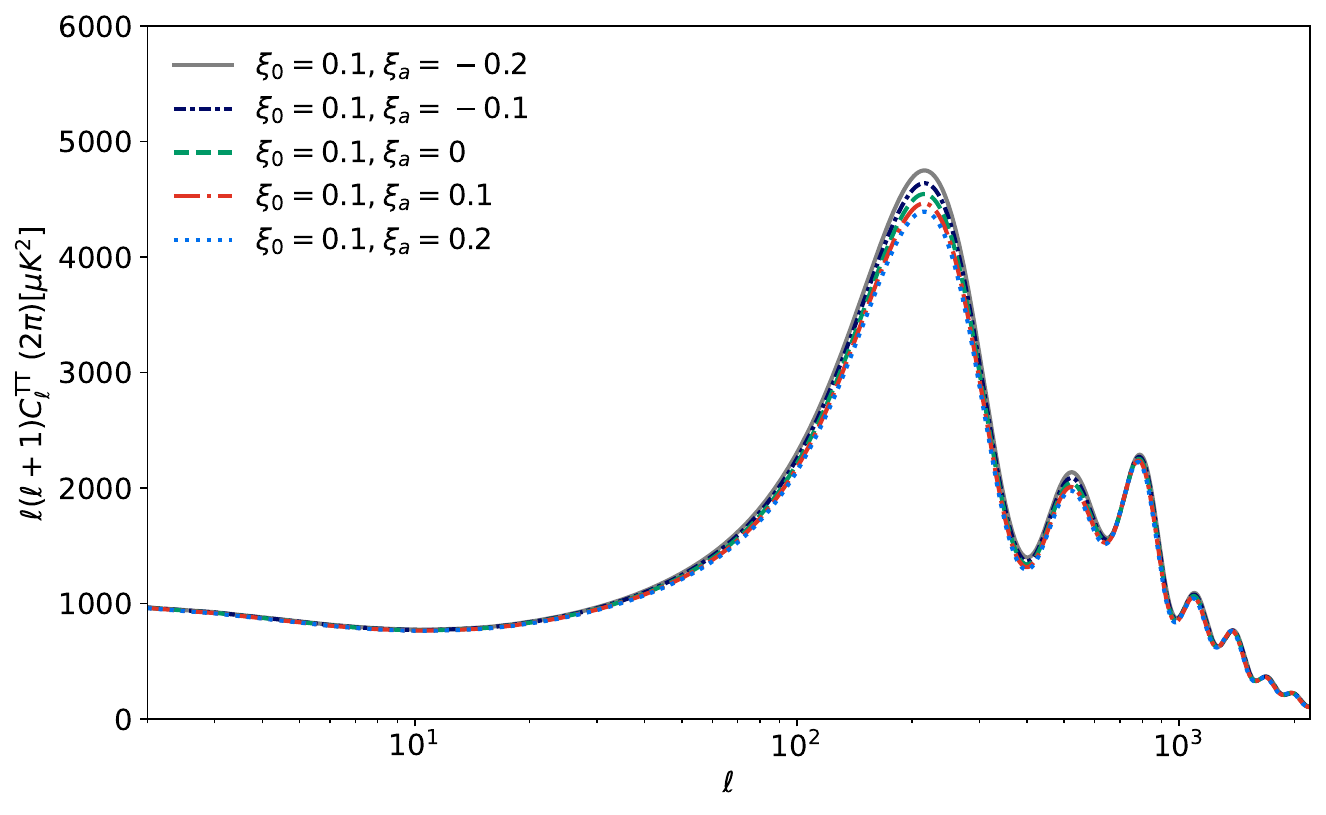}
    \includegraphics[width=0.45\textwidth]{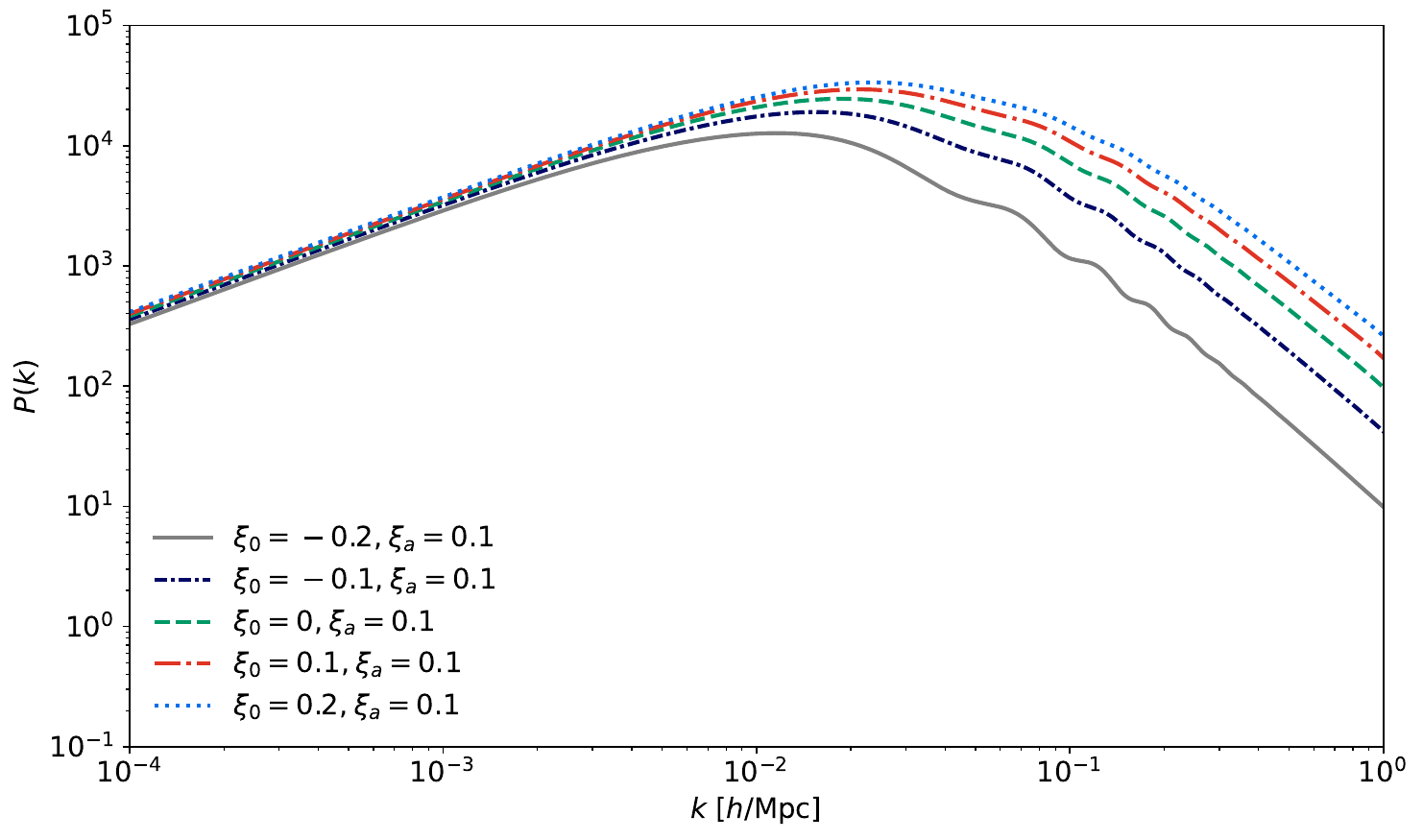}
    \includegraphics[width=0.45\textwidth]{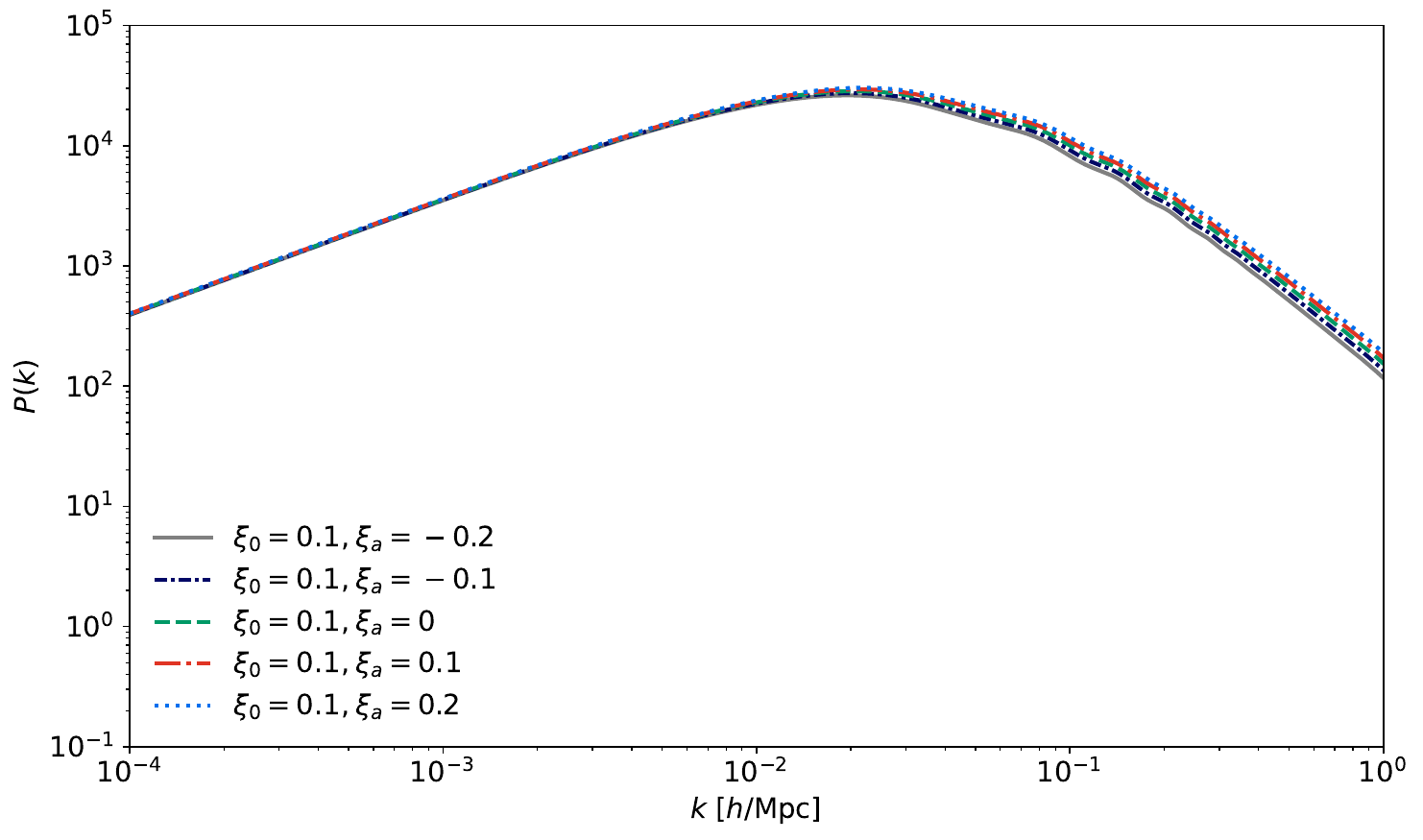}
    \caption{In the upper panel, we show the CMB TT spectra for {\bf IVS1a}, considering two separate cases: the left graph corresponds to fixed $\xi_a$ with varying values of $\xi_0$, while the right graph corresponds to fixed $\xi_0$ with varying $\xi_a$. In the lower panel, we show the matter power spectra for the same two cases: the left graph corresponds to fixed $\xi_a$ with varying $\xi_0$, and the right graph corresponds to fixed $\xi_0$ with varying $\xi_a$. The other parameters needed for drawing these plots have been fixed as usual, and here we took the constraints from CMB+DESI+PantheonPlus as described in the next section~\ref{sec-results}.  }
    \label{fig:cmb+matter-power-IVS1a}
\end{figure*}

\begin{figure*}
    \centering
    \includegraphics[width=0.45\textwidth]{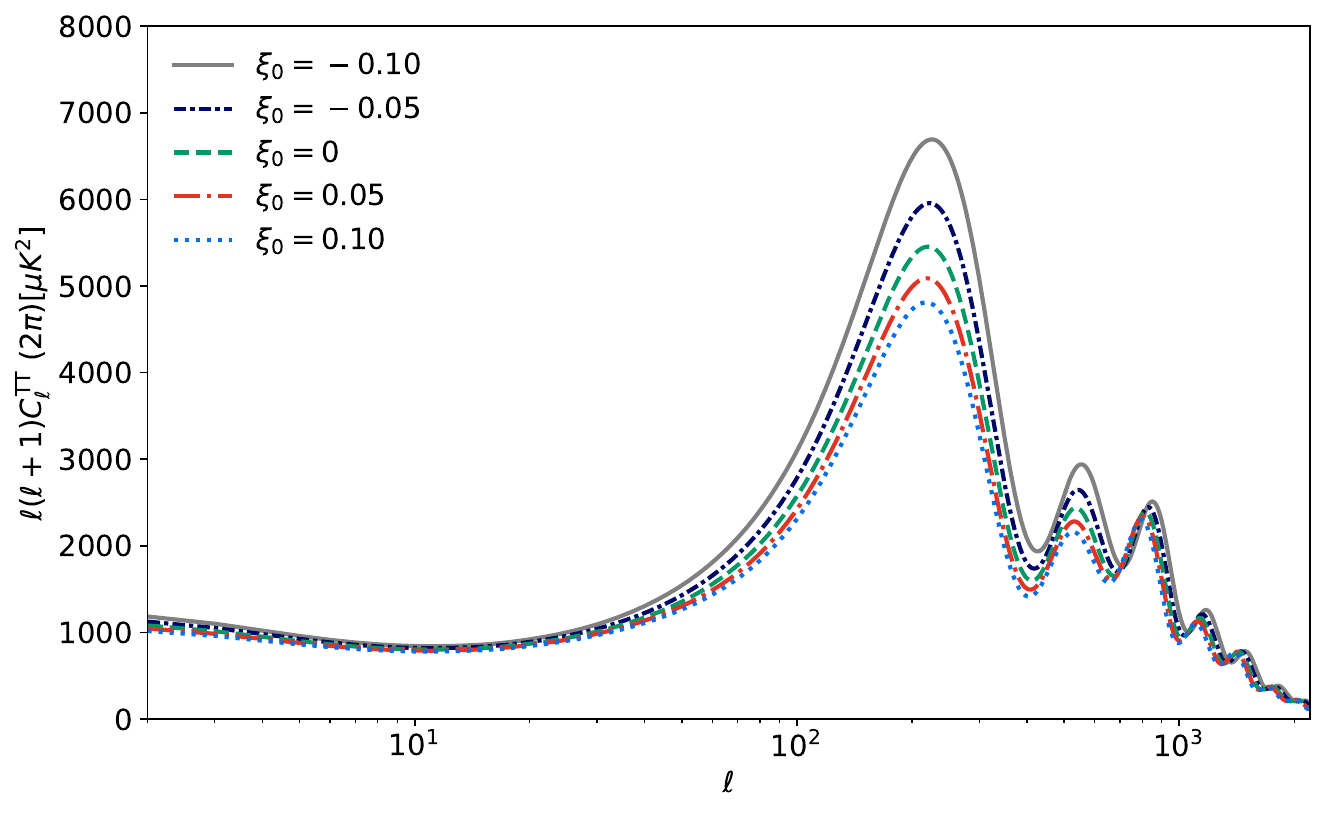}
    \includegraphics[width=0.45\textwidth]{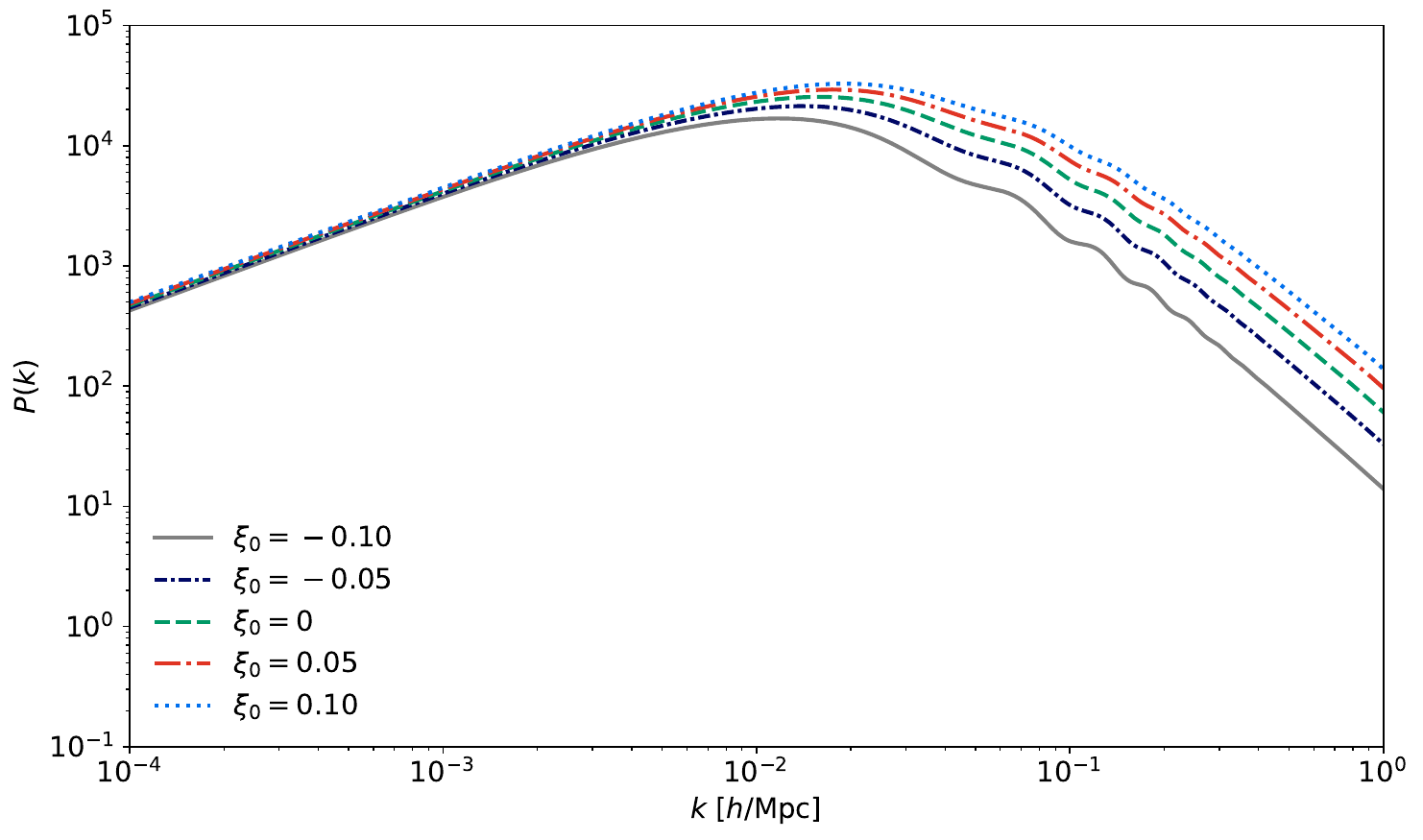}
\caption{We show the CMB TT spectra (left plot) and the matter power spectra (right plot) for {\bf IVS1b}, considering various values of $\xi_0$.   The other parameters needed for drawing these plots have been fixed as usual, and here we took the constraints from CMB+DESI+PantheonPlus as described in the next section~\ref{sec-results}. }
    \label{fig:cmb+matter-power-IVS1b}
\end{figure*}

\begin{figure*}
    \centering
    \includegraphics[width=0.45\textwidth]{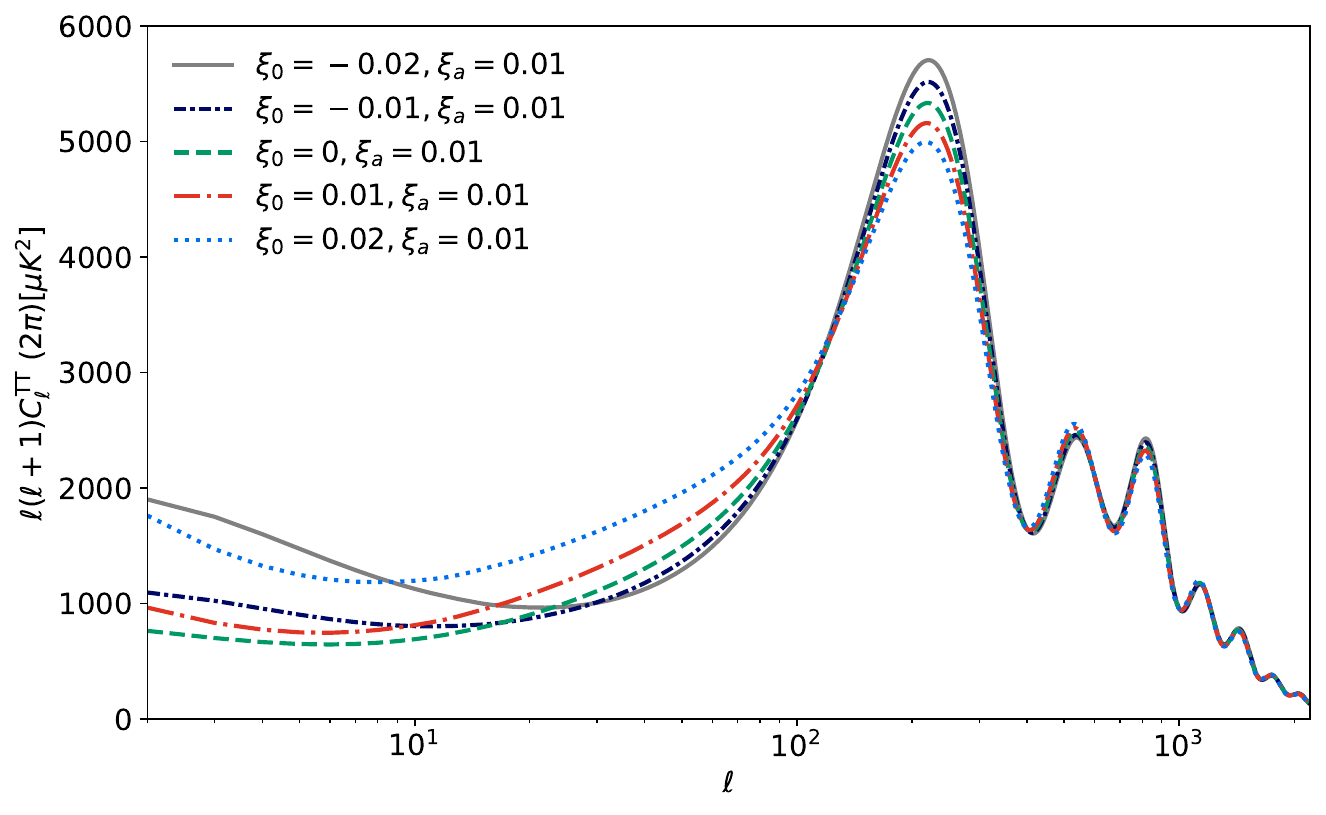}
    \includegraphics[width=0.45\textwidth]{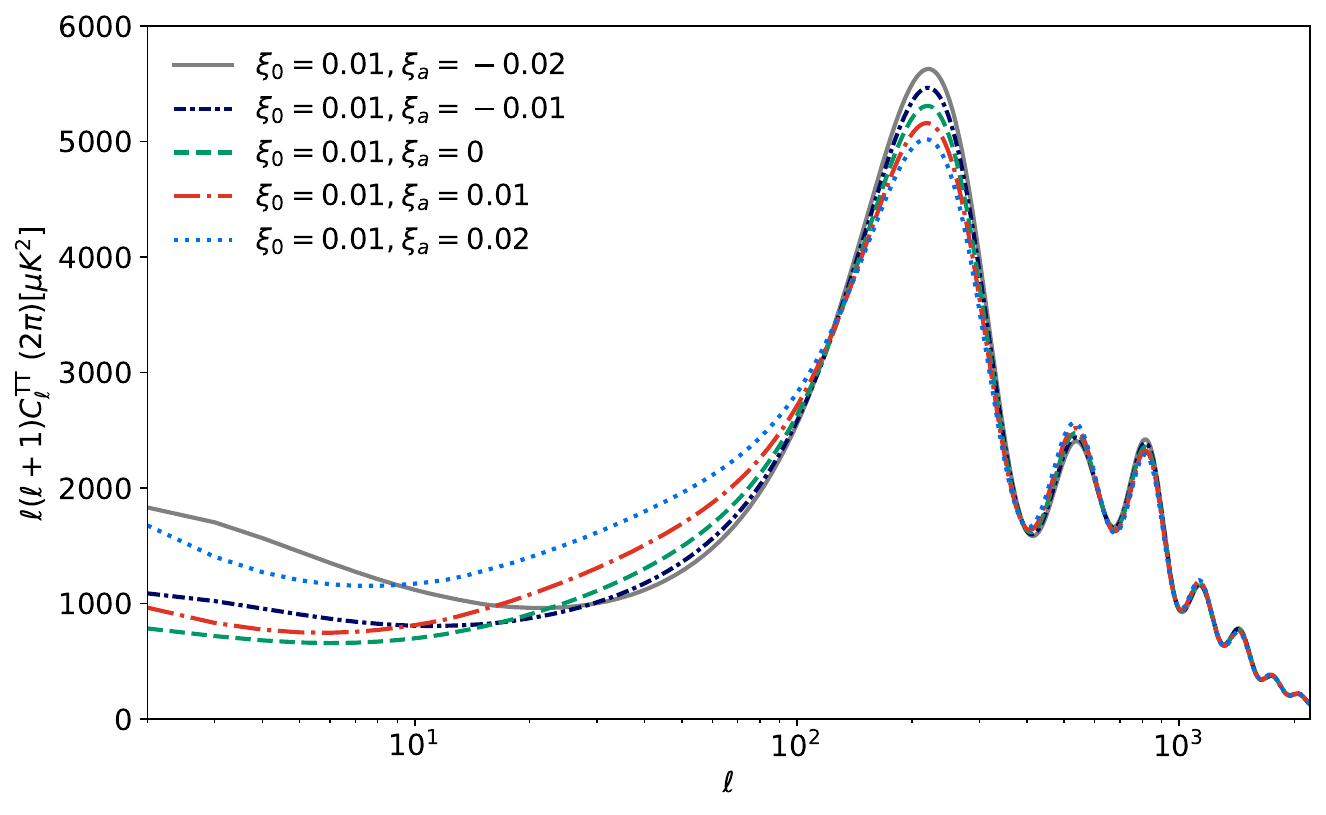}
    \includegraphics[width=0.45\textwidth]{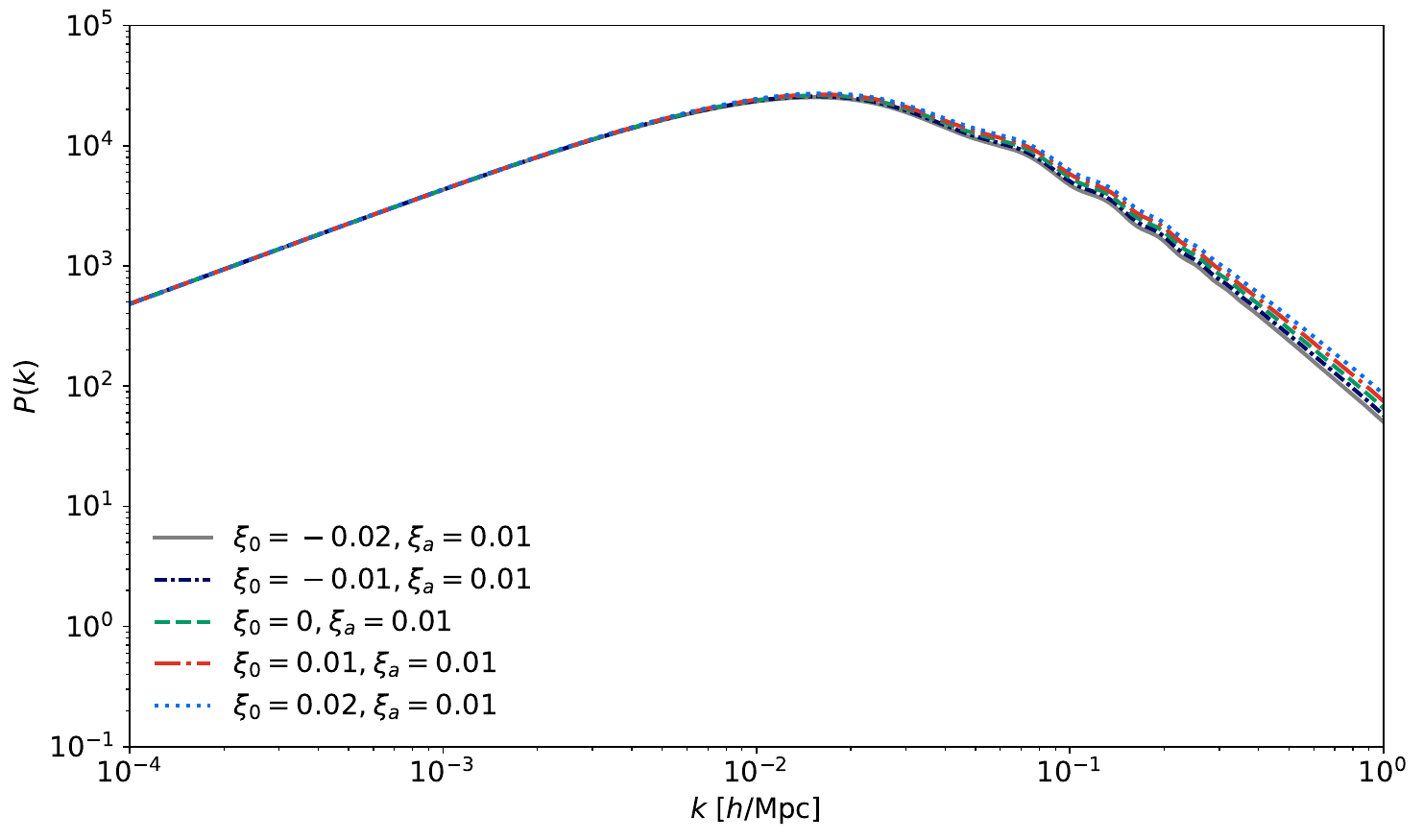}
    \includegraphics[width=0.45\textwidth]{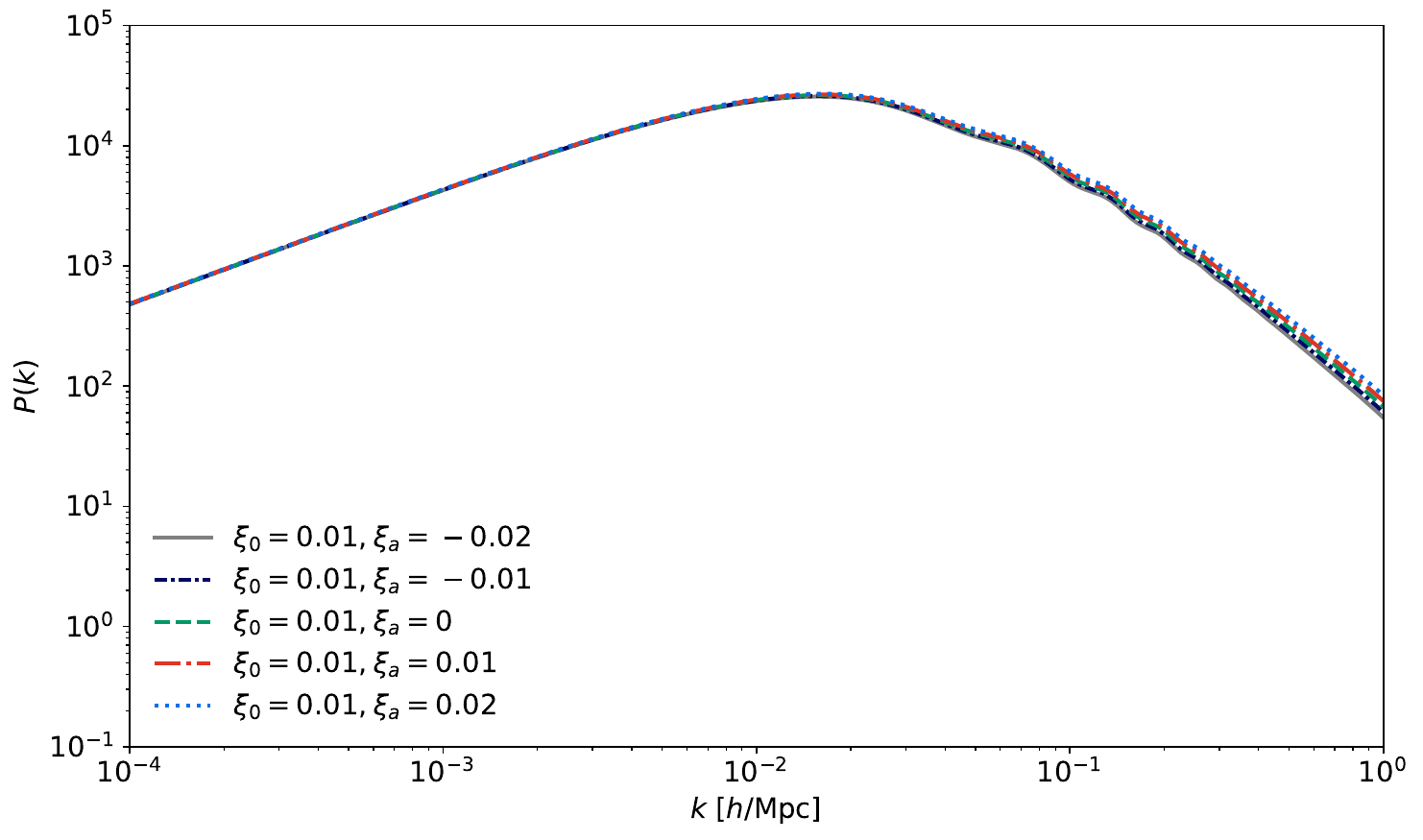}
\caption{In the upper panel, we show the CMB TT spectra for {\bf IVS2a}, considering two separate cases: the left graph corresponds to fixed $\xi_a$ with varying values of $\xi_0$, while the right graph corresponds to fixed $\xi_0$ with varying $\xi_a$. In the lower panel, we show the matter power spectra for the same two cases: the left graph corresponds to fixed $\xi_a$ with varying $\xi_0$, and the right graph corresponds to fixed $\xi_0$ with varying $\xi_a$.  The other parameters needed for drawing these plots have been fixed as usual, and here we took the constraints from CMB+DESI+PantheonPlus as described in the next section~\ref{sec-results}.  }
    \label{fig:cmb+matter-power-IVS2a}
\end{figure*}
\begin{figure*}
    \centering
    \includegraphics[width=0.45\textwidth]{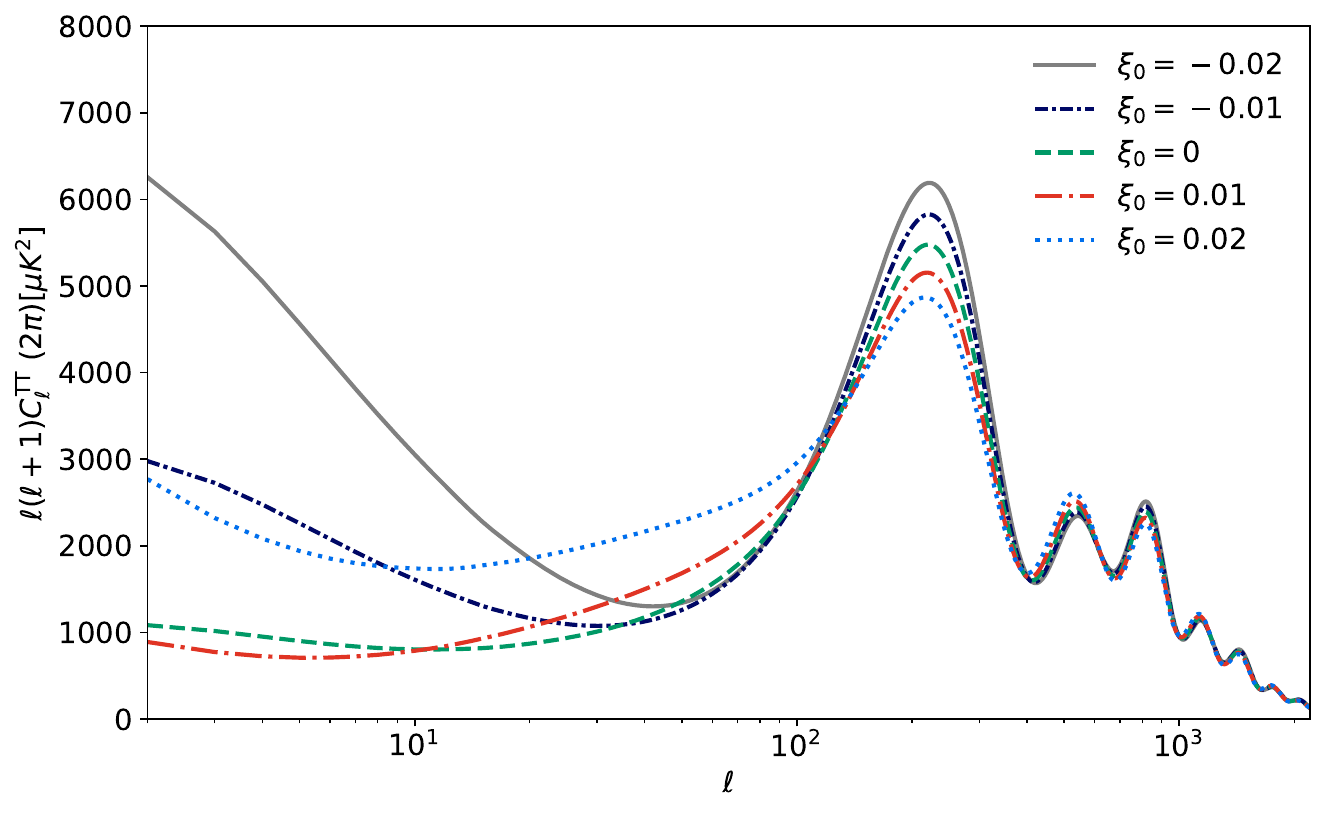}
    \includegraphics[width=0.45\textwidth]{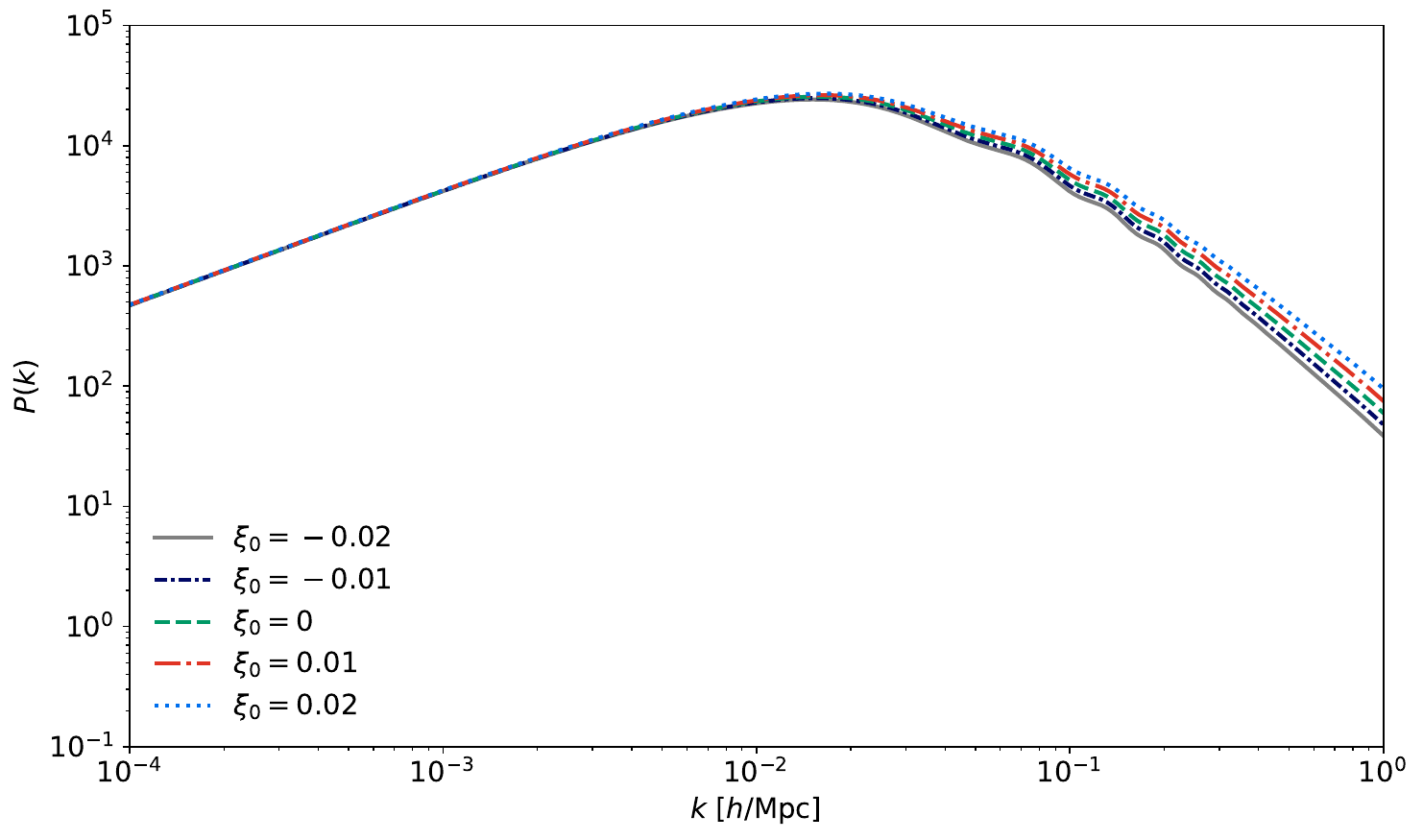}
    \caption{We show the CMB TT spectra (left plot) and the matter power spectra (right plot) for {\bf IVS2b}, considering various values of $\xi_0$. The other parameters needed for drawing these plots have been fixed as usual, and here we took the constraints from CMB+DESI+PantheonPlus as described in the next section~\ref{sec-results}.  }
    \label{fig:cmb+matter-power-IVS2b}
\end{figure*}

Finally, in Figs.~\ref{fig:cmb+matter-power-IVS1a}, \ref{fig:cmb+matter-power-IVS1b}, \ref{fig:cmb+matter-power-IVS2a}, and \ref{fig:cmb+matter-power-IVS2b}, we show the influence of the interacting scenarios on the CMB TT power spectra and matter power spectra, exploring different values of the free parameters in the coupling functions.

In the upper left panel of Fig.~\ref{fig:cmb+matter-power-IVS1a}, we show the CMB TT spectra for {\bf IVS1a}, considering a fixed value of $\xi_a$ and various negative and positive values of $\xi_0$. It is clear from this plot that for different present-day values of $\xi(a)$, the CMB TT spectrum is affected in both the low- and high-multipole regions, with more noticeable changes appearing at high multipoles. In particular, as $\xi_0$ increases, the height of the first acoustic peak of the TT spectrum increases.
A similar feature is observed in the lower left panel of Fig.~\ref{fig:cmb+matter-power-IVS1a}, which shows the corresponding effects on the matter power spectrum for the same parameter values. In the upper right and lower right panels of Fig.~\ref{fig:cmb+matter-power-IVS1a}, we show the CMB TT spectra and matter power spectra, respectively, for a fixed value of $\xi_0$ while varying $\xi_a$ over positive and negative values. In contrast to the previous case, here we do not observe any notable changes in either of the two observables.
In the left and right panels of Fig.~\ref{fig:cmb+matter-power-IVS1b}, we show the influence on the CMB TT and matter power spectra, respectively, for {\bf IVS1b}, considering various positive and negative values of $\xi_0$. We find that varying $\xi_0$ does not lead to any significant changes in either the CMB TT spectrum or the matter power spectrum.

In the upper left panel of Fig.~\ref{fig:cmb+matter-power-IVS2a}, we show the CMB TT spectra for {\bf IVS2a}, again considering a fixed value of $\xi_a$ and various negative and positive values of $\xi_0$. In the upper right panel of Fig.~\ref{fig:cmb+matter-power-IVS2a}, we show the CMB TT spectra for a fixed value of $\xi_0$ and different values of $\xi_a$. From both plots, one can clearly see that in the low-$\ell$ region, significant changes appear in the CMB TT spectra. These changes are associated with the late-time integrated Sachs–Wolfe effect. 
In the high-$\ell$ region, although the effects of interaction are also visible, they are not as pronounced compared to those at low multipoles. In contrast, for the matter power spectra in this scenario (see the lower panels of Fig.~\ref{fig:cmb+matter-power-IVS2a}), we do not observe any notable changes.
Fig.~\ref{fig:cmb+matter-power-IVS2b} corresponds to {\bf IVS2b}, where we show the CMB TT spectra (left panel) and matter power spectra (right panel) for different values of $\xi_0$. Once again, we observe significant changes in the low-$\ell$ region, which reflect the late-time integrated Sachs–Wolfe effect. Additionally, we notice some mild changes in the matter power spectra, caused by the interaction strength as quantified by $\xi_0$.

\section{Observational datasets and methodology}
\label{sec-data}

In this section, we present the observational datasets and the statistical methodology adopted to constrain the interacting scenarios considered in this work. We begin by describing the observational datasets:

\begin{itemize}
		
\item {\bf Cosmic Microwave Background (CMB)} measurements from the Planck 2018 release~\cite{Planck:2018vyg,Planck:2019nip}, incorporating temperature and polarization angular power spectra through the {\it plikTTTEEE+lowl+lowE} likelihood combination.
		
\item {\bf Baryon Acoustic Oscillation (BAO)} cosmic distance measurements derived from the second data release of the {\bf DESI DR2} survey~\cite{DESI:2025zgx}.
		
\item {\bf Type Ia Supernovae} observations from three major compilations: (i) the {\bf PantheonPlus} sample~\cite{Scolnic:2021amr}; (ii) the 2087-object {\bf Union3} catalog~\cite{Rubin:2023ovl}; and (iii) the five-year supernova dataset from the {\bf Dark Energy Survey} ({\bf DESY5})~\cite{DES:2024jxu}.
		
\end{itemize} 

We now turn to the statistical methodology adopted in this work. For the interacting models considered, we modified the publicly available cosmological code {\tt CAMB}~\cite{Lewis:1999bs,Howlett:2012mh}, and performed Markov Chain Monte Carlo (MCMC) analyses using the publicly available sampler {\tt Cobaya} ({\bf co}de for {\bf bay}esian {\bf a}nalysis)~\cite{Torrado:2020dgo}. The convergence of the MCMC chains is tested using the Gelman–Rubin parameter $R - 1$~\cite{Gelman:1992zz}, and we continue running the chains until the condition $R - 1 < 0.03$ is satisfied. 
In Table~\ref{tab:priors}, we display the flat priors on the free parameters of the interacting scenarios.

Finally, we perform Bayesian model comparison by computing the logarithm of the Bayesian evidence, $\ln \mathcal{Z}$, using \texttt{MCEvidence}~\citep{Heavens:2017afc}. For a model $\mathcal{M}_i$ with parameters $\Theta$, the evidence is
\begin{equation}
\mathcal{Z}_i = \int \mathcal{L}(D | \Theta, \mathcal{M}_i)\, \pi(\Theta | \mathcal{M}_i)\, \mathrm{d}\Theta,
\end{equation}
where $\mathcal{L}$ is the likelihood and $\pi$ the prior. Model comparison is based on the Bayes factor,
\begin{equation}
\ln B_{ij} = \ln \mathcal{Z}_i - \ln \mathcal{Z}_j,
\end{equation}
where $\ln B_{ij} > 0$ favors model $\mathcal{M}_i$ over $\mathcal{M}_j$.

We interpret $\ln B_{ij}$ using the revised Jeffreys’ scale~\citep{Kass:1995loi}: \textit{inconclusive} ($0\!-\!1$), \textit{weak} ($1\!-\!2.5$), \textit{moderate} ($2.5\!-\!5$), \textit{strong} ($5\!-\!10$), and \textit{very strong} ($>10$) evidence.

\begin{table*}
	\begin{center}
		\renewcommand{\arraystretch}{1.4}
		\begin{tabular}{|c@{\hspace{1 cm}}|@{\hspace{1 cm}}c|@{\hspace{1 cm}}c|@{\hspace{1 cm}}c|@{\hspace{1 cm}}c|}
			\hline
			\textbf{Parameter}   & \textbf{Prior (IVS1a)} & \textbf{Prior (IVS2a)} & \textbf{Prior (IVS1b)} & \textbf{Prior (IVS2b)} \\
			\hline\hline
			$\Omega_{b} h^2$     &$[0.005,0.1]$ &$[0.005,0.1]$ &$[0.005,0.1]$ &$[0.005,0.1]$\\
			$\Omega_{c} h^2$     &$[0.01,0.99]$ &$[0.01,0.99]$ &$[0.01,0.99]$ &$[0.01,0.99]$\\
			$\tau$               &$[0.01,0.8]$  &$[0.01,0.8]$  &$[0.01,0.8]$ &$[0.01,0.8]$\\
			$n_s$                &$[0.8, 1.2]$  &$[0.8, 1.2]$  &$[0.8, 1.2]$ &$[0.8, 1.2]$\\
			$\log[10^{10}A_{s}]$ &$[1.61,3.91]$ &$[1.61,3.91]$ &$[1.61,3.91]$ &$[1.61,3.91]$\\
			$100\theta_{MC}$     &$[0.5,10]$    &$[0.5,10]$    &$[0.5,10]$ &$[0.5,10]$ \\ 
			$\xi_0$              &$[-1,1]$      &$[-1,1]$      &$[-1,1]$  &$[-1,1]$ \\
			$\xi_a$              &$[-1,1]$     &$[-1,1]$      &$--$       &$--$\\
				
			\hline
		\end{tabular}
	\end{center}
	\caption{Flat priors on the free parameters of the interacting scenarios used in the statistical analysis.}
	\label{tab:priors}
\end{table*}
\begingroup
\squeezetable                                   
\begin{center}  
	\begin{table*}
		\begin{tabular}{cccccc}
			\hline\hline
			Parameters & CMB & CMB+DESI & CMB+DESI+PantheonPlus & CMB+DESI+Union3 & CMB+DESI+DESY5 \\ \hline
			
		$\Omega_b h^2$ & 
		$0.02231_{-0.00015-0.00029}^{+0.00015+0.00030}$ &
		$0.02245_{-0.00014-0.00027}^{+0.00014+0.00028}$ &
		$0.02246_{-0.00014-0.00027}^{+0.00014+0.00027}$ &
		$0.00246_{-0.00014-0.00027}^{+0.00014+0.00027}$ &
		$0.02246_{-0.00014-0.00026}^{+0.00014+0.00027}$ \\

        $\Omega_c h^2$ & 
		$<0.09527<0.14097$ &
		$0.11693_{-0.00822-0.05259}^{+0.02961+0.03335}$ &
		$0.13743_{-0.00329-0.01668}^{+0.01027+0.01159}$ &
		$0.13804_{-0.00281-0.01728}^{+0.00987+0.01095}$ &
		$0.14178_{-0.00184-0.01140}^{+0.00623+0.00709}$ \\

		$100\theta_{MC}$ & 
		$1.0439_{-0.0039-0.0048}^{+0.0025+0.0053}$ &
		$1.04102_{-0.0017-0.0022}^{+0.00053+0.0032}$ &
		$1.03985_{-0.00061-0.00092}^{+0.00036+0.0011}$ &
		$1.03983_{-0.00061-0.00092}^{+0.00033+0.0011}$ &
		$1.03965_{-0.00045-0.00075}^{+0.00031+0.00083}$ \\
		
		$\tau$ & 
		$0.0540_{-0.0080-0.015}^{+0.0080+0.016}$ &
		$0.0577_{-0.0084-0.016}^{+0.0084+0.017}$ &
		$0.0581_{-0.0083-0.015}^{+0.0075+0.017}$ &
		$0.0578_{-0.0085-0.015}^{+0.0074+0.017}$ &
		$0.0581_{-0.0083-0.015}^{+0.0073+0.017}$ \\
		
		$n_s$ & 
		$0.9723_{-0.0043-0.0087}^{+0.0043+0.0082}$ &
		$0.9771_{-0.0037-0.0074}^{+0.0037+0.0073}$ &
		$0.9774_{-0.0037-0.0073}^{+0.0037+0.0074}$ &
		$0.9774_{-0.0037-0.0073}^{+0.0037+0.0073}$ &
		$0.9777_{-0.0038-0.0072}^{+0.0037+0.0072}$ \\
		
		${\rm{ln}}(10^{10} A_s)$ & 
		$3.055_{-0.016-0.031}^{+0.016+0.033}$ &
		$3.057_{-0.017-0.033}^{+0.017+0.035}$ &
		$3.058_{-0.017-0.031}^{+0.017+0.034}$ &
		$3.057_{-0.017-0.032}^{+0.015+0.035}$ &
		$3.057_{-0.017-0.032}^{+0.017+0.034}$ \\

		$\xi_0$ & 
		$0.19_{-0.17-0.34}^{+0.20+0.32}$ &
		$-0.051_{-0.16-0.22}^{+0.075+0.28}$ &
		$-0.146_{-0.074-0.11}^{+0.045+0.13}$ &
		$-0.147_{-0.071-0.11}^{+0.041+0.13}$ &
		$-0.159_{-0.052-0.084}^{+0.034+0.098}$ \\
		
		$\xi_a$ & 
		$<0.0186$ &
		$0.21_{-0.22-0.59}^{+0.32+0.51}$ &
		$0.34_{-0.16-0.35}^{+0.18+0.32}$ &
		$0.34_{-0.16-0.38}^{+0.18+0.32}$ &
		$0.33_{-0.14-0.29}^{+0.14+0.28}$ \\	
		
		$\Omega_m$ & 
		$0.207_{-0.15-0.17}^{+0.070+0.22}$ &
		$0.295_{-0.025-0.085}^{+0.070+0.084}$ &
		$0.347_{-0.011-0.041}^{+0.024+0.032}$ &
		$0.349_{-0.0099-0.032}^{+0.025+0.044}$ &
		$0.361_{-0.0070-0.023}^{+0.016+0.030}$ \\
		
		$\sigma_8$ & 
		$1.58_{-0.93-0.98}^{+0.16+2.2}$ &
		$0.8583_{-0.00050-0.21}^{+0.19+0.41}$ &
		$0.713_{-0.044-0.053}^{+0.014+0.077}$ &
		$0.710_{-0.043-0.052}^{+0.013+0.080}$ &
		$0.6929_{-0.027-0.035}^{+0.0096+0.050}$ \\
		
		$H_0$ [Km/s/Mpc] & 
		$69.2_{-4.2-9}^{+4.9+8}$ &
		$69.07_{-1.1-1.9}^{+0.75+2.1}$ &
		$68.06_{-0.54-1.1}^{+0.54+1.1}$ &
		$67.98_{-0.59-1.1}^{+0.59+1.2}$ &
		$67.60_{-0.49-0.95}^{+0.49+0.95}$ \\
		
		$S_8$ & 
		$1.074_{-0.28-0.33}^{+0.091+0.54}$ &
		$0.8175_{-0.076-0.094}^{+0.0092+0.16}$ &
		$0.765_{-0.022-0.033}^{+0.012+0.039}$ &
		$0.764_{-0.022-0.033}^{+0.011+0.040}$ &
		$0.759_{-0.016-0.026}^{+0.010+0.029}$ \\
		
		$r_{\rm{drag}}$ [Mpc] & 
		$147.11_{-0.29-0.58}^{+0.29+0.57}$ &
		$147.49_{-0.24-0.46}^{+0.24+0.48}$ &
		$147.52_{-0.23-0.46}^{+0.23+0.46}$ &
		$147.52_{-0.24-0.46}^{+0.24+0.48}$ &
		$147.55_{-0.24-0.46}^{+0.24+0.48}$ \\
		\hline 
		$\rm{ln}\mathcal{B}_{ij}$ & 
		$-2.9$ & 
		$-5.5$ & 
		$-5.8$ & 
		$-5.5$ & 
		$-4.0$ \\				
		\hline\hline                                                         
	\end{tabular}                                                       
	\caption{68\% and 95\% CL constraints on the cosmological parameters of the {\bf IVS1a} scenario for the CMB, CMB+DESI, CMB+DESI+PantheonPlus, CMB+DESI+Union3, and CMB+DESI+DESY5 data combinations.}
	\label{table-IVS1a}                   
\end{table*}                                     
\end{center}
\endgroup

\begingroup
\squeezetable                                  
\begin{center}  
	\begin{table*}
		\begin{tabular}{cccccc}
			\hline\hline
			Parameters & CMB & CMB+DESI & CMB+DESI+PantheonPlus & CMB+DESI+Union3 & CMB+DESI+DESY5 \\ \hline
			
			$\Omega_b h^2$ & 
			$0.02236_{-0.00016-0.00030}^{+0.00016+0.00031}$ &
			$0.02247_{-0.00013-0.00026}^{+0.00013+0.00025}$ &
			$0.02251_{-0.00013-0.00026}^{+0.00013+0.00026}$ &
			$0.02250_{-0.00013-0.00026}^{+0.00013+0.00026}$ &
			$0.02253_{-0.00013-0.00025}^{+0.00013+0.00025}$ \\

            $\Omega_c h^2$ & 
			$0.11921_{-0.00143-0.00280}^{+0.00140+0.00290}$ &
			$0.11742_{-0.00086-0.00173}^{+0.00085+0.00170}$ &
			$0.11673_{-0.00082-0.00163}^{+0.00081+0.00161}$ &
			$0.11680_{-0.00085-0.00165}^{+0.00085+0.00165}$ &
			$0.11642_{-0.00081-0.00162}^{+0.00082+0.00163}$ \\
			
			$100\theta_{MC}$ & 
			$1.04071_{-0.00032-0.00062}^{+0.00032+0.00063}$ &
			$1.04108_{-0.00029-0.00057}^{+0.00029+0.00057}$ &
			$1.04103_{-0.00028-0.00056}^{+0.00028+0.00056}$ &
			$1.04103_{-0.00028-0.00054}^{+0.00028+0.00055}$ &
			$1.04107_{-0.00028-0.00056}^{+0.00028+0.00056}$ \\
			
			$\tau$ & 
			$0.0532_{-0.0080-0.016}^{+0.0080+0.016}$ &
			$0.0573_{-0.0081-0.016}^{+0.0081+0.017}$ &
			$0.0592_{-0.0088-0.015}^{+0.0075+0.017}$ &
			$0.0590_{-0.0085-0.015}^{+0.0075+0.017}$ &
			$0.0600_{-0.0087-0.015}^{+0.0075+0.017}$ \\
			
			$n_s$ & 
			$0.9740_{-0.0046-0.0091}^{+0.0046+0.0089}$ &
			$0.9793_{-0.0034-0.0068}^{+0.0034+0.0067}$ &
			$0.9791_{-0.0034-0.0066}^{+0.0034+0.0067}$ &
			$0.9789_{-0.0034-0.0066}^{+0.0034+0.0068}$ &
			$0.9797_{-0.0035-0.0069}^{+0.0035+0.0067}$ \\
			
			${\rm{ln}}(10^{10} A_s)$ & 
			$3.052_{-0.017-0.033}^{+0.017+0.033}$ &
			$3.055_{-0.017-0.033}^{+0.017+0.034}$ &
			$3.058_{-0.018-0.032}^{+0.016+0.035}$ &
			$3.058_{-0.017-0.032}^{+0.016+0.034}$ &
			$3.059_{-0.017-0.032}^{+0.017+0.035}$ \\
			
			$\xi_0$ & 
			$0.109_{-0.10-0.18}^{+0.092+0.20}$ &
			$0.048_{-0.031-0.059}^{+0.031+0.062}$ &
			$-0.002_{-0.021-0.040}^{+0.021+0.040}$ &
			$0.002_{-0.024-0.045}^{+0.024+0.048}$ &
			$-0.023_{-0.019-0.037}^{+0.019+0.038}$ \\
			
			$\Omega_m$ & 
			$0.287_{-0.033-0.056}^{+0.028+0.063}$ &
			$0.2912_{-0.0046-0.0088}^{+0.0046+0.0092}$ &
			$0.3010_{-0.0049-0.0095}^{+0.0049+0.0097}$ &
			$0.3002_{-0.0055-0.011}^{+0.0055+0.011}$ &
			$0.3055_{-0.0047-0.0091}^{+0.0047+0.0093}$ \\
			
			$\sigma_8$ & 
			$1.06_{-0.30-0.40}^{+0.11+0.55}$ &
			$0.829_{-0.014-0.027}^{+0.016+0.029}$ &
			$0.803_{-0.036-0.064}^{+0.032+0.070}$ &
			$0.810_{-0.045-0.077}^{+0.037+0.081}$ &
			$0.769_{-0.032-0.057}^{+0.029+0.062}$ \\
			
			$H_0$ [km/s/Mpc] & 
			$70.7_{-3.9-6.6}^{+3.3+7.5}$ &
			$68.95_{-0.48-0.96}^{+0.48+0.90}$ &
			$68.18_{-0.56-1.1}^{+0.56+1.1}$ &
			$68.29_{-0.65-1.2}^{+0.65+1.3}$ &
			$67.60_{-0.52-1.0}^{+0.52+1.0}$ \\
			
			$S_8$ & 
			$1.024_{-0.23-0.30}^{+0.082+0.41}$ &
			$0.816_{-0.014-0.027}^{+0.014+0.027}$ &
			$0.804_{-0.032-0.059}^{+0.029+0.063}$ &
			$0.810_{-0.040-0.068}^{+0.033+0.071}$ &
			$0.776_{-0.028-0.053}^{+0.028+0.058}$ \\
			
			$r_{\rm{drag}}$ [Mpc] & 
			$147.17_{-0.30-0.60}^{+0.30+0.60}$ &
			$147.52_{-0.22-0.44}^{+0.22+0.44}$ &
			$147.66_{-0.22-0.43}^{+0.22+0.43}$ &
			$147.65_{-0.22-0.44}^{+0.22+0.43}$ &
			$147.72_{-0.22-0.42}^{+0.22+0.43}$ \\
			\hline 
			$\rm{ln}\mathcal{B}_{ij}$ & 
			$-2.5$ & 
			$-4.2$ & 
			$-5.8$ & 
			$-5.6$ & 
			$-5.1$ \\			
			\hline\hline                                                         
		\end{tabular}                                                       
		\caption{68\% and 95\% CL constraints on the cosmological parameters of the {\bf IVS1b} scenario for the CMB, CMB+DESI, CMB+DESI+PantheonPlus, CMB+DESI+Union3, and CMB+DESI+DESY5 data combinations. }
		\label{table-IVS1b}                   
	\end{table*}                                     
\end{center}
\endgroup

\section{Results}
\label{sec-results}

\begin{figure*}
	\includegraphics[width=0.8\textwidth]{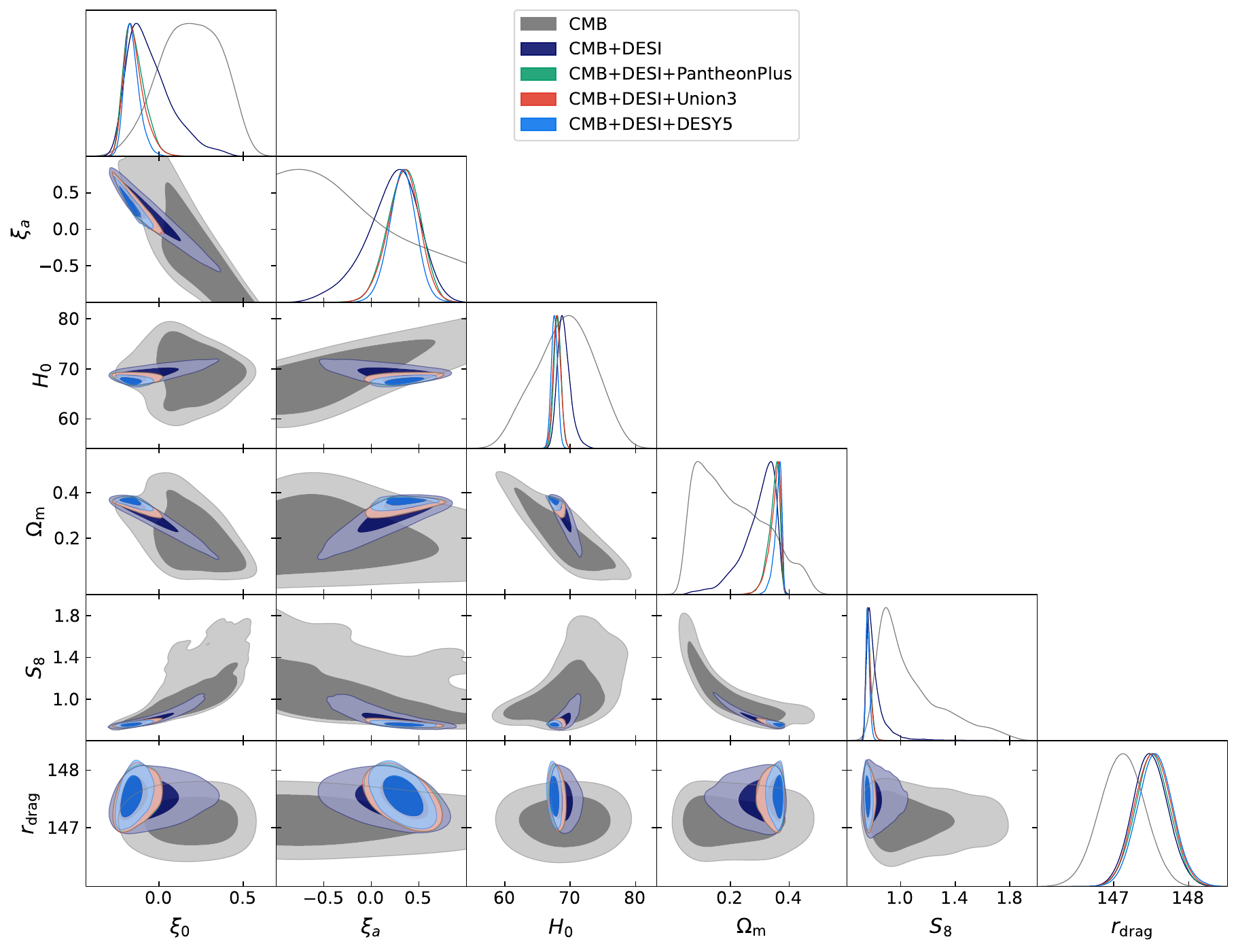}
	\caption{One-dimensional posterior distributions and two-dimensional joint contours for the most relevant parameters of the {\bf IVS1a} scenario, using different combinations of cosmological measurements.}
	\label{fig:IVS1a}
\end{figure*}

\begin{figure*}
	\includegraphics[width=0.8\textwidth]{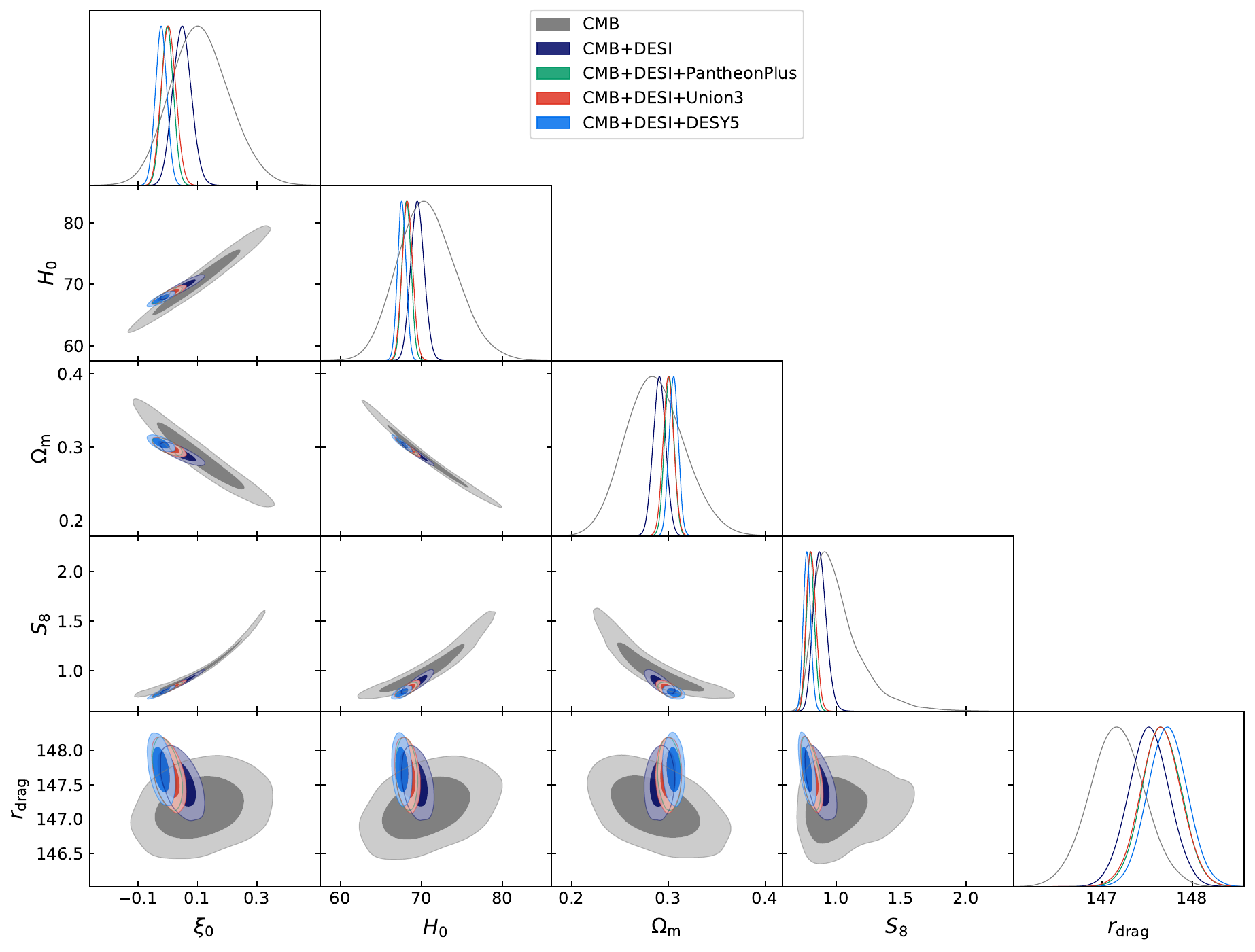}
	\caption{One-dimensional posterior distributions and two-dimensional joint contours for the most relevant parameters of the {\bf IVS1b} scenario, using several combinations of cosmological datasets.}
	\label{fig:IVS1b}
\end{figure*}

\begin{figure*}
    \centering
    \includegraphics[width=0.45\textwidth]{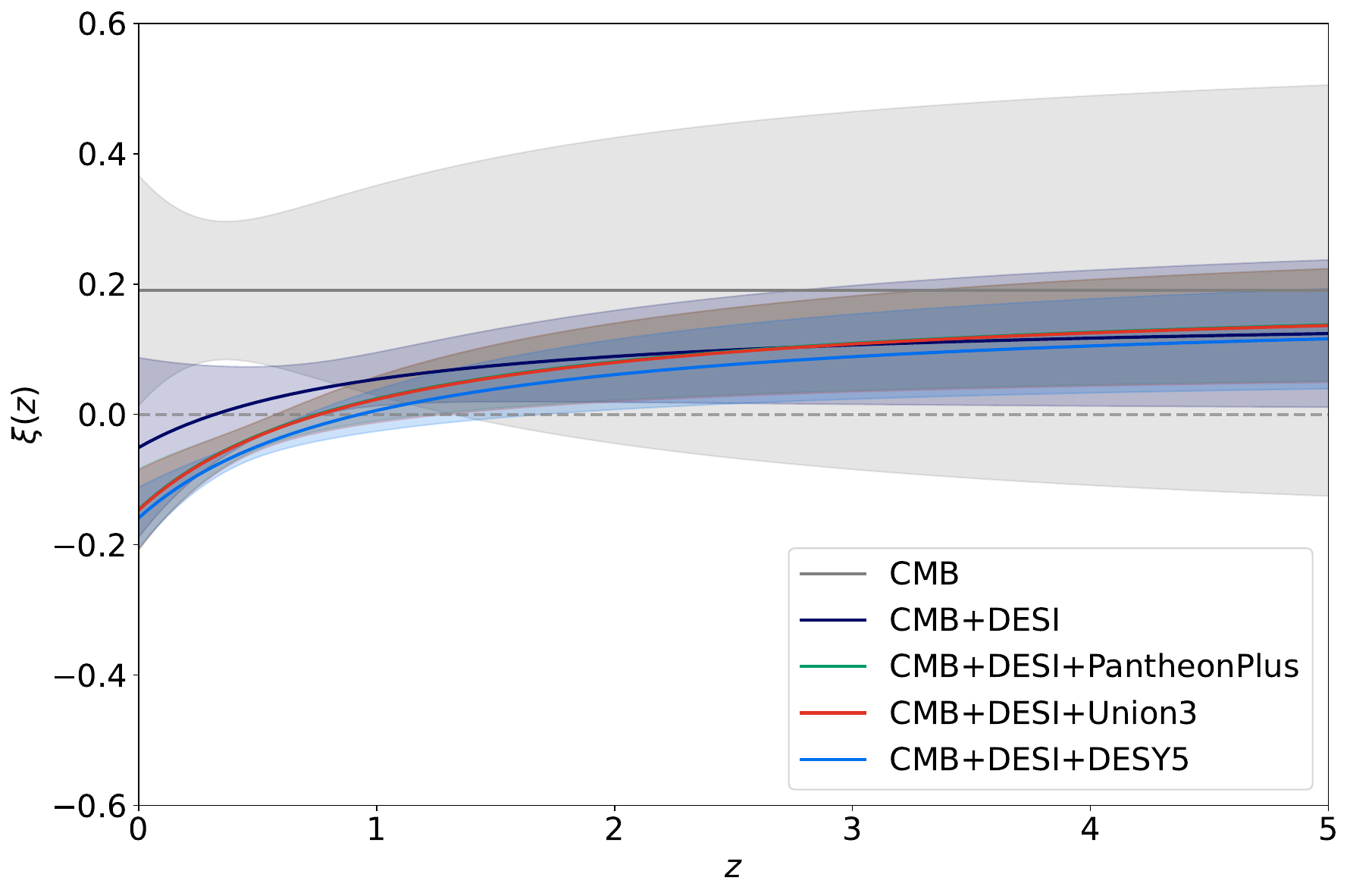}
    \includegraphics[width=0.45\textwidth]{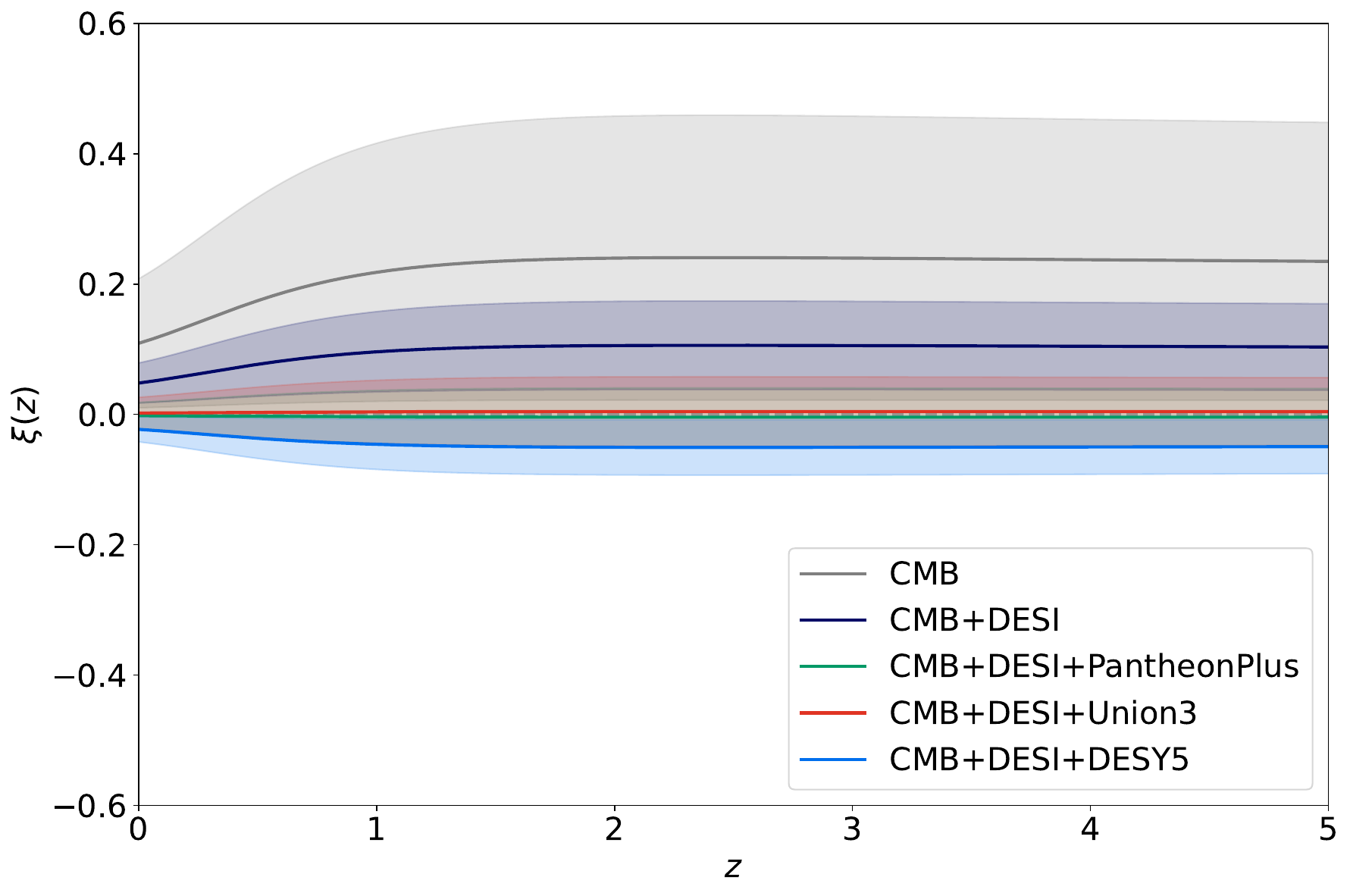}
    \caption{Evolution of $\xi(z)$ ($1 + z = a^{-1}$) with its 68\% CL region for {\bf IVS1a} (left plot) and {\bf IVS1b} (right plot), considering all the datasets. The horizontal dotted line in each plot corresponds to the no-interaction case (i.e., $\xi(z) = 0$). Note that in the right plot, the dotted line is not clearly visible.}
    \label{fig:xi(z)-ivs1a1b}
\end{figure*}

In this section, we present the constraints on the four interacting scenarios, considering the various astronomical datasets summarized in Section~\ref{sec-data}. For each interacting scenario, we begin by constraining it using CMB data alone. We then combine the CMB with BAO measurements from DESI DR2 (i.e., CMB+DESI), and subsequently add three different SNIa samples—PantheonPlus, Union3, and DESY5—to the CMB+DESI combination. Overall, we perform five different analyses.
In Tables~\ref{table-IVS1a}, \ref{table-IVS1b}, \ref{table-IVS2a}, and \ref{table-IVS2b}, we summarize the constraints on the interacting scenarios, and in Figs.~\ref{fig:IVS1a}, \ref{fig:IVS1b}, \ref{fig:xi(z)-ivs1a1b}, \ref{fig:IVS2a}, \ref{fig:IVS2b}, and \ref{fig:xi(z)-ivs2a2b}, we show the corresponding graphical representations.
In what follows, we describe the constraints on each individual interacting scenario.

\subsection{IVS1a}

The constraints on {\bf IVS1a} are summarized in Table~\ref{table-IVS1a} and Fig.~\ref{fig:IVS1a}. Note that for {\bf IVS1a}, the coupling function $\xi(a)$ has two free parameters, namely $\xi_0$ and $\xi_a$. We begin with the constraints from CMB alone and then progressively describe the impact of including additional datasets.

For the CMB alone, we find $\xi_0 \neq 0$ at slightly more than 68\% CL, while $\xi_a$ is only weakly constrained and reaches an upper bound. This indicates evidence for interaction, although no conclusive statement can be made regarding the dynamical nature of the coupling parameter. The estimated value of the Hubble constant is higher, with large error bars: $H_0 = 69.2^{+4.9}_{-4.2}$ km/s/Mpc at 68\% CL. The $H_0$ tension is therefore significantly alleviated due to both the higher mean value and the enlarged uncertainty. Since the present-day value of the coupling is positive ($\xi_0 = 0.19^{+0.20}_{-0.17}$ at 68\% CL), this suggests an energy flow from CDM to DE, resulting in a lower value of the matter density parameter: $\Omega_m = 0.207^{+0.070}_{-0.15}$ at 68\% CL.

When DESI data are added to the CMB (CMB+DESI), notable changes appear. Most significantly, $\xi_0$ changes sign—from positive (in the CMB-only case) to negative—and $\xi_a$ becomes constrained. Both parameters are now consistent with their null values within 68\% CL. However, the negative mean value of $\xi_0$ ($\xi_0 = -0.051^{+0.075}_{-0.16}$ at 68\% CL) indicates an energy flow from DE to CDM, leading to an increased value of $\Omega_m$ compared to the CMB-only case. The Hubble constant keeps taking relatively large values, $H_0 = 69.07^{+0.75}_{-1.1}$ km/s/Mpc at 68\% CL, but with significantly reduced error bars, further alleviating the $H_0$ tension. 
As can also be seen from Fig.~\ref{fig:IVS1a}, in the CMB-only case, the correlation between $\Omega_m$ and $H_0$ is rather weak, in contrast to the standard $\Lambda$CDM scenario where the two parameters are tightly anti-correlated. The addition of DESI significantly strengthens the correlations between $\Omega_m$ and the coupling parameters $\xi_0$ and $\xi_a$, while at the same time it reduces the correlation between $\Omega_m$ and $H_0$. This explains why, despite the increase in $\Omega_m$, the value of $H_0$ remains higher than in the standard $\Lambda$CDM model.

The results become particularly interesting when SNIa data are added to CMB+DESI. For both CMB+DESI+PantheonPlus and CMB+DESI+Union3, we find $\xi_0 < 0$ at more than 95\% CL, and $\xi_a \neq 0$ at more than 68\% CL. This suggests not only strong evidence for an interaction but also for a dynamical coupling function. The non-zero nature of $\xi_a$ is further strengthened when using CMB+DESI+DESY5, where it is found to be non-zero at more than 95\% CL. In all three SNIa combinations, the negative value of $\xi_0$ (indicating energy flow from DE to CDM) corresponds to higher values of $\Omega_m$ compared to CMB alone.
The constraints on the Hubble constant remain similar across the three SNIa combinations. Interestingly, we observe significantly lower values of the $S_8$ parameter in all three cases: $S_8 = 0.765^{+0.012}_{-0.022}$ at 68\% CL (CMB+DESI+PantheonPlus), $0.764^{+0.011}_{-0.022}$ at 68\% CL (CMB+DESI+Union3), and $0.759^{+0.010}_{-0.016}$ at 68\% CL (CMB+DESI+DESY5), all substantially lower than the Planck-$\Lambda$CDM prediction~\cite{Planck:2018vyg}.

However, despite these phenomenological improvements, Bayesian evidence analysis shows that the $\Lambda$CDM scenario remains favored over this interacting model. This may be attributed to the two additional free parameters in {\bf IVS1a} compared to the standard $\Lambda$CDM model.

\subsection{IVS1b}

We now focus on the {\bf IVS1b} scenario, in which the coupling parameter $\xi(a)$ is dynamical but depends on only one free parameter, namely $\xi_0$ (see Eq.~(\ref{xi(a)-2})). The constraints on this interacting scenario are summarized in Table~\ref{table-IVS1b} and Fig.~\ref{fig:IVS1b}. 

Considering the constraints from CMB alone, we find marginal evidence for interaction, with $\xi_0 = 0.109^{+0.092}_{-0.10}$ at 68\% CL, along with a relatively high value of the Hubble constant, $H_0 = 70.7^{+3.3}_{-3.9}$ km/s/Mpc at 68\% CL. This alleviates the $H_0$ tension quite effectively. The marginal preference for interaction also leads to a slightly lower value of $\Omega_m$.
Contrary to the previous case ({\bf IVS1a}), here the correlation between $\Omega_m$ and $H_0$ remains very strong, which explains the systematic shifts observed when additional data are included. Moreover, the coupling parameter $\xi_0$ is negatively correlated with $\Omega_m$ and positively correlated with $H_0$. As a result, even a small indication of $\xi_0 \neq 0$ leads to opposite shifts in $\Omega_m$ and $H_0$: when $\xi_0 > 0$, $\Omega_m$ tends to decrease and $H_0$ increases, while for $\xi_0 < 0$, the opposite trend is observed.

When DESI is combined with CMB (CMB+DESI), the evidence for interaction becomes more pronounced. In this case, we find $\xi_0 > 0$ at slightly more than 68\% CL ($\xi_0 = 0.048^{+0.031}_{-0.031}$ at 68\% CL). The Hubble constant is slightly reduced compared to the CMB-only case, with $H_0 = 68.95^{+0.48}_{-0.48}$ km/s/Mpc at 68\% CL, while the value of $S_8$ ($= 0.816^{+0.014}_{-0.014}$ at 68\% CL) is consistent with the Planck-$\Lambda$CDM prediction~\cite{Planck:2018vyg}.

We now consider the inclusion of three different SNIa samples with CMB+DESI. According to the results, for both CMB+DESI+PantheonPlus and CMB+DESI+Union3, there is no significant evidence for a non-zero $\xi_0$. In particular, for these two cases, the mean value of $\xi_0$ is very close to zero ($\xi_0 \sim 0.002$), and its 68\% CL interval includes the null value. The corresponding $H_0$ values are slightly increased (around $H_0 \sim 68$ km/s/Mpc), and the $S_8$ values are slightly reduced compared to the Planck-$\Lambda$CDM estimate~\cite{Planck:2018vyg}.

However, for CMB+DESI+DESY5, we find evidence for interaction at more than 68\% CL, with $\xi_0 = -0.023^{+0.019}_{-0.019}$ at 68\% CL. Despite this, none of the five combinations statistically favor the interacting model over $\Lambda$CDM. According to the Bayesian evidence analysis, the standard $\Lambda$CDM model remains preferred.

We close this section with Fig.~\ref{fig:xi(z)-ivs1a1b}, where we show the time evolution of the coupling parameters defined in Eqs.~(\ref{xi(a)-1}) and~(\ref{xi(a)-2}), including their 68\% CL regions, considering all the datasets. The left plot in this figure illustrates a pronounced time variation in $\xi(z)$, particularly at late times, corresponding to the two-parameter model. In contrast, the right plot, which corresponds to the one-parameter $\xi(z)$ model, shows a much slower evolution of the coupling function. Overall, both plots indicate a clear tendency toward interaction in the dark sector.

\begingroup
\squeezetable                                  
\begin{center}  
	\begin{table*}
		\begin{tabular}{cccccc}
			\hline\hline
			Parameters & CMB & CMB+DESI & CMB+DESI+PantheonPlus & CMB+DESI+Union3 & CMB+DESI+DESY5 \\ \hline
			
			$\Omega_b h^2$ & 
			$0.02241_{-0.00016-0.00032}^{+0.00016+0.00031}$ &
			$0.02257_{-0.00015-0.00030}^{+0.00015+0.00030}$ &
			$0.02256_{-0.00015-0.00029}^{+0.00015+0.00030}$ &
			$0.02256_{-0.00015-0.00029}^{+0.00015+0.00029}$ &
			$0.02255_{-0.00015-0.00031}^{+0.00015+0.00030}$ \\

            $\Omega_c h^2$ & 
			$0.11826_{-0.00151-0.00295}^{+0.00151+0.00299}$ &
			$0.11520_{-0.00066-0.00132}^{+0.00066+0.00132}$ &
			$0.11530_{-0.00068-0.00133}^{+0.00069+0.00138}$ &
			$0.11528_{-0.00069-0.00138}^{+0.00070+0.00135}$ &
			$0.11527_{-0.00069-0.00137}^{+0.00069+0.00137}$ \\
			
			$100\theta_{MC}$ & 
			$1.04068_{-0.00033-0.00065}^{+0.00033+0.00064}$ &
			$1.04108_{-0.00028-0.00055}^{+0.00028+0.00054}$ &
			$1.04109_{-0.00028-0.00053}^{+0.00028+0.00055}$ &
			$1.04110_{-0.00028-0.00056}^{+0.00028+0.00055}$ &
			$1.04108_{-0.00028-0.00056}^{+0.00028+0.00053}$ \\
			
			$\tau$ & 
			$0.0535_{-0.0078-0.015}^{+0.0078+0.015}$ &
			$0.0571_{-0.0082-0.016}^{+0.0082+0.016}$ &
			$0.0578_{-0.0079-0.015}^{+0.0079+0.016}$ &
			$0.0574_{-0.0088-0.015}^{+0.0075+0.017}$ &
			$0.0582_{-0.0084-0.015}^{+0.0072+0.017}$ \\
			
			$n_s$ & 
			$0.9715_{-0.0049-0.0095}^{+0.0049+0.0095}$ &
			$0.9793_{-0.0033-0.0066}^{+0.0033+0.0064}$ &
			$0.9788_{-0.0034-0.0068}^{+0.0034+0.0067}$ &
			$0.9787_{-0.0033-0.0065}^{+0.0033+0.0065}$ &
			$0.9781_{-0.0034-0.0066}^{+0.0034+0.0068}$ \\
			
			${\rm{ln}}(10^{10} A_s)$ & 
			$3.055_{-0.016-0.032}^{+0.016+0.032}$ &
			$3.055_{-0.017-0.033}^{+0.017+0.034}$ &
			$3.057_{-0.016-0.032}^{+0.016+0.033}$ &
			$3.055_{-0.018-0.032}^{+0.016+0.035}$ &
			$3.057_{-0.018-0.031}^{+0.015+0.035}$ \\
			
			$\xi_0$ & 
			$0.030_{-0.049-0.13}^{+0.071+0.12}$ &
			$0.075_{-0.039-0.089}^{+0.048+0.084}$ &
			$0.033_{-0.041-0.089}^{+0.047+0.087}$ &
			$0.039_{-0.042-0.094}^{+0.051+0.090}$ &
			$-0.009_{-0.046-0.095}^{+0.051+0.093}$ \\
			
			$\xi_a$ & 
			$-0.028_{-0.071-0.12}^{+0.049+0.13}$ &
			$-0.075_{-0.048-0.084}^{+0.039+0.089}$ &
			$-0.033_{-0.047-0.087}^{+0.041+0.090}$ &
			$-0.039_{-0.051-0.091}^{+0.042+0.094}$ &
			$0.009_{-0.048-0.093}^{+0.048+0.095}$ \\
			
			$\Omega_m$ & 
			$0.314_{-0.013-0.024}^{+0.012+0.026}$ &
			$0.2912_{-0.0046-0.0088}^{+0.0046+0.0092}$ &
			$0.2954_{-0.0045-0.0087}^{+0.0045+0.0091}$ &
			$0.2947_{-0.0046-0.0089}^{+0.0046+0.0092}$ &
			$0.2989_{-0.0044-0.0084}^{+0.0044+0.0089}$ \\
			
			$\sigma_8$ & 
			$0.825_{-0.016-0.038}^{+0.020+0.035}$ &
			$0.829_{-0.014-0.032}^{+0.016+0.029}$ &
			$0.817_{-0.015-0.031}^{+0.015+0.029}$ &
			$0.818_{-0.015-0.033}^{+0.017+0.029}$ &
			$0.805_{-0.016-0.031}^{+0.016+0.030}$ \\
			
			$H_0$ [km/s/Mpc] & 
			$67.1_{-1.1-2.2}^{+1.1+2.0}$ &
			$68.95_{-0.48-0.96}^{+0.48+0.90}$ &
			$68.48_{-0.47-0.95}^{+0.47+0.92}$ &
			$68.56_{-0.48-0.94}^{+0.48+0.92}$ &
			$68.06_{-0.46-0.93}^{+0.46+0.87}$ \\
			
			$S_8$ & 
			$0.844_{-0.018-0.036}^{+0.018+0.036}$ &
			$0.816_{-0.014-0.027}^{+0.014+0.027}$ &
			$0.811_{-0.014-0.028}^{+0.014+0.027}$ &
			$0.811_{-0.014-0.029}^{+0.014+0.027}$ &
			$0.804_{-0.015-0.028}^{+0.015+0.028}$ \\
			
			$r_{\rm{drag}}$ [Mpc] & 
			$146.90_{-0.37-0.71}^{+0.37+0.73}$ &
			$147.54_{-0.23-0.46}^{+0.23+0.46}$ &
			$147.52_{-0.23-0.46}^{+0.23+0.45}$ &
			$147.53_{-0.24-0.46}^{+0.24+0.45}$ &
			$147.54_{-0.24-0.47}^{+0.24+0.47}$ \\
			\hline 
			$\rm{ln}\mathcal{B}_{ij}$ & 
			$-9.7$ & 
			$-10.9$ & 
			$-12.1$ & 
			$-11.8$ & 
			$-12.4$ \\			
			\hline\hline                                                         
		\end{tabular}                                                       
		\caption{68\% and 95\% CL constraints on the cosmological parameters of the {\bf IVS2a} scenario for the CMB, CMB+DESI, CMB+DESI+PantheonPlus, CMB+DESI+Union3, and CMB+DESI+DESY5 data combinations. }
		\label{table-IVS2a}                   
	\end{table*}                                     
\end{center}
\endgroup
	
\begin{figure*}
	\includegraphics[width=0.8\textwidth]{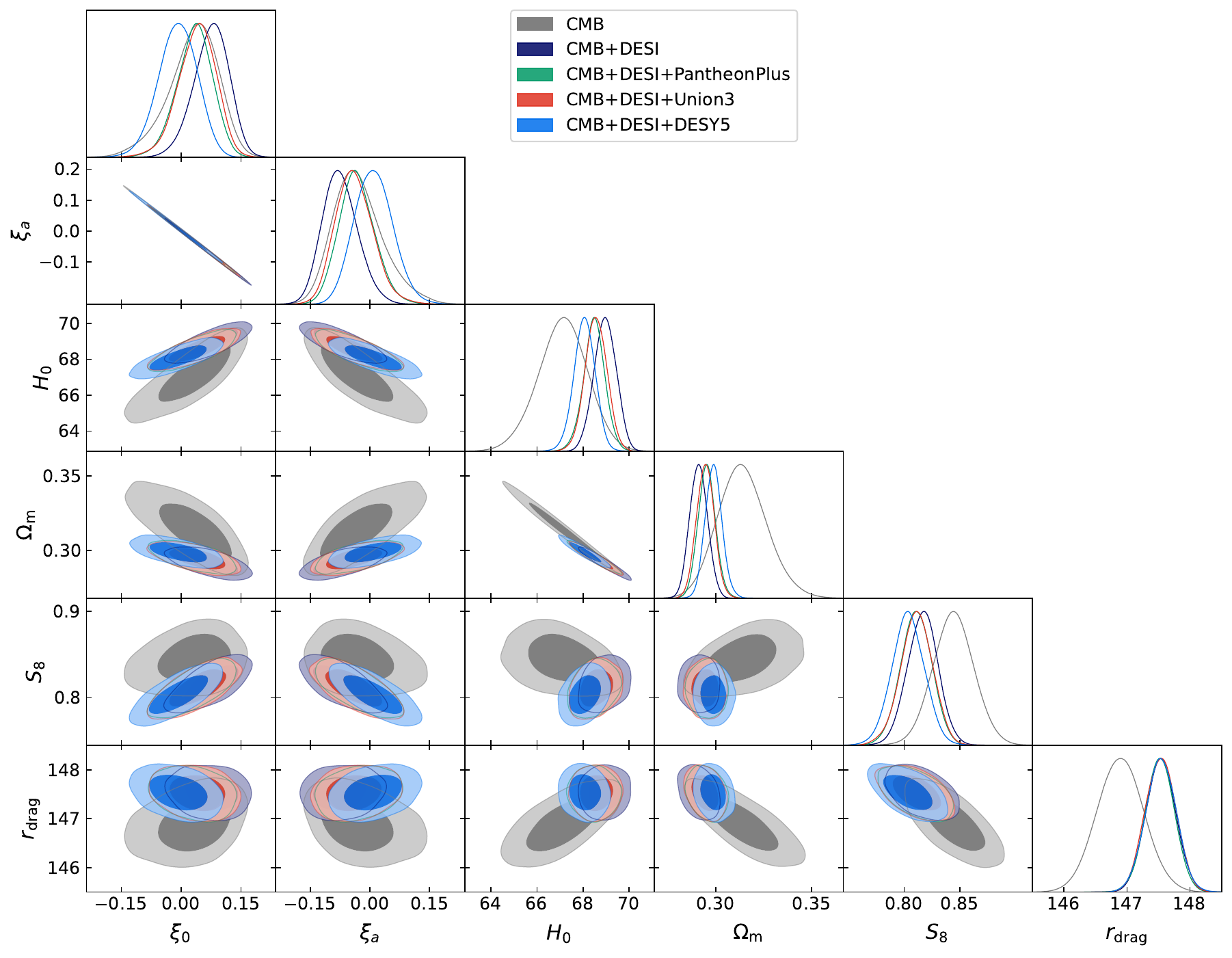}
	\caption{One-dimensional posterior distributions and two-dimensional joint contours for the most relevant parameters of the {\bf IVS2a} scenario, using different combinations of cosmological measurements.}
	\label{fig:IVS2a}
\end{figure*}

\begingroup
\squeezetable                                  
\begin{center}  
	\begin{table*}
		\begin{tabular}{cccccc}
			\hline\hline
			Parameters & CMB & CMB+DESI & CMB+DESI+PantheonPlus & CMB+DESI+Union3 & CMB+DESI+DESY5 \\ \hline
			
			$\Omega_b h^2$ & 
			$0.02240_{-0.00016-0.00031}^{+0.00016+0.00031}$ &
			$0.02258_{-0.00015-0.00030}^{+0.00015+0.00031}$ &
			$0.02257_{-0.00016-0.00030}^{+0.00016+0.00031}$ &
			$0.02257_{-0.00015-0.00029}^{+0.00015+0.00030}$ &
			$0.02255_{-0.00015-0.00029}^{+0.00015+0.00030}$ \\

            $\Omega_c h^2$ & 
			$0.11932_{-0.00142-0.00278}^{+0.00140+0.00281}$ &
			$0.11569_{-0.00068-0.00131}^{+0.00069+0.00133}$ &
			$0.11590_{-0.00066-0.00127}^{+0.00064+0.00129}$ &
			$0.11586_{-0.00067-0.00126}^{+0.00062+0.00131}$ &
			$0.11612_{-0.00065-0.00132}^{+0.00065+0.00130}$ \\
			
			$100\theta_{MC}$ & 
			$1.04062_{-0.00032-0.00062}^{+0.00032+0.00062}$ &
			$1.04108_{-0.00028-0.00055}^{+0.00028+0.00055}$ &
			$1.04106_{-0.00028-0.00054}^{+0.00028+0.00055}$ &
			$1.04106_{-0.00028-0.00055}^{+0.00028+0.00055}$ &
			$1.04105_{-0.00028-0.00055}^{+0.00028+0.00056}$ \\
			
			$\tau$ & 
			$0.0538_{-0.0079-0.015}^{+0.0079+0.016}$ &
			$0.0591_{-0.0082-0.015}^{+0.0073+0.016}$ &
			$0.0589_{-0.0086-0.015}^{+0.0074+0.017}$ &
			$0.0592_{-0.0086-0.015}^{+0.0073+0.018}$ &
			$0.0586_{-0.0083-0.015}^{+0.0074+0.017}$ \\
			
			$n_s$ & 
			$0.9704_{-0.0044-0.0087}^{+0.0044+0.0086}$ &
			$0.9792_{-0.0034-0.0067}^{+0.0034+0.0066}$ &
			$0.9786_{-0.0034-0.0066}^{+0.0034+0.0066}$ &
			$0.9787_{-0.0033-0.0066}^{+0.0033+0.0063}$ &
			$0.9781_{-0.0033-0.0065}^{+0.0033+0.0066}$ \\
			
			${\rm{ln}}(10^{10} A_s)$ & 
			$3.057_{-0.016-0.032}^{+0.016+0.032}$ &
			$3.058_{-0.017-0.032}^{+0.016+0.035}$ &
			$3.058_{-0.018-0.032}^{+0.016+0.035}$ &
			$3.059_{-0.018-0.032}^{+0.016+0.036}$ &
			$3.058_{-0.017-0.032}^{+0.017+0.035}$ \\
			
			$\xi_0$ & 
			$0.00079_{-0.00055-0.0011}^{+0.00055+0.0011}$ &
			$0.00036_{-0.00056-0.0011}^{+0.00056+0.0011}$ &
			$0.00037_{-0.00056-0.0011}^{+0.00056+0.0011}$ &
			$0.00036_{-0.00055-0.0011}^{+0.00055+0.0011}$ &
			$0.00039_{-0.00055-0.0011}^{+0.00055+0.0011}$ \\
			
			$\Omega_m$ & 
			$0.3202_{-0.0089-0.017}^{+0.0089+0.018}$ &
			$0.2978_{-0.0039-0.0074}^{+0.0039+0.0077}$ &
			$0.2991_{-0.0037-0.0072}^{+0.0037+0.0073}$ &
			$0.2989_{-0.0037-0.0072}^{+0.0037+0.0076}$ &
			$0.3005_{-0.0038-0.0074}^{+0.0038+0.0075}$ \\
			
			$\sigma_8$ & 
			$0.8162_{-0.0082-0.016}^{+0.0082+0.016}$ &
			$0.8051_{-0.0074-0.014}^{+0.0074+0.015}$ &
			$0.8058_{-0.0077-0.014}^{+0.0077+0.016}$ &
			$0.8059_{-0.0080-0.014}^{+0.0070+0.016}$ &
			$0.8066_{-0.0076-0.014}^{+0.0076+0.015}$ \\
			
			$H_0$ [km/s/Mpc] & 
			$66.70_{-0.62-1.2}^{+0.62+1.2}$ &
			$68.29_{-0.31-0.60}^{+0.31+0.60}$ &
			$68.20_{-0.30-0.58}^{+0.30+0.57}$ &
			$68.21_{-0.30-0.61}^{+0.30+0.59}$ &
			$68.10_{-0.30-0.58}^{+0.30+0.59}$ \\
			
			$S_8$ & 
			$0.843_{-0.017-0.034}^{+0.017+0.034}$ &
			$0.802_{-0.010-0.019}^{+0.010+0.020}$ &
			$0.805_{-0.010-0.019}^{+0.010+0.021}$ &
			$0.804_{-0.010-0.019}^{+0.010+0.021}$ &
			$0.807_{-0.010-0.019}^{+0.010+0.020}$ \\
			
			$r_{\rm{drag}}$ [Mpc] & 
			$146.83_{-0.34-0.67}^{+0.34+0.68}$ &
			$147.61_{-0.24-0.46}^{+0.24+0.48}$ &
			$147.57_{-0.24-0.46}^{+0.24+0.47}$ &
			$147.58_{-0.23-0.47}^{+0.23+0.45}$ &
			$147.52_{-0.24-0.47}^{+0.24+0.47}$ \\
			\hline 
            
			$\rm{ln}\mathcal{B}_{ij}$ & 
			$-7.2$ & 
			$-9.2$ & 
			$-9.1$ & 
			$-9.0$ & 
			$-8.8$ \\			
			\hline\hline                                                         
		\end{tabular}                                                       
		\caption{68\% and 95\% CL constraints on the cosmological parameters of the {\bf IVS2b} scenario for the CMB, CMB+DESI, CMB+DESI+PantheonPlus, CMB+DESI+Union3, and CMB+DESI+DESY5 data combinations. }
		\label{table-IVS2b}                   
	\end{table*}                                     
\end{center}
\endgroup

\begin{figure*}
	\includegraphics[width=0.8\textwidth]{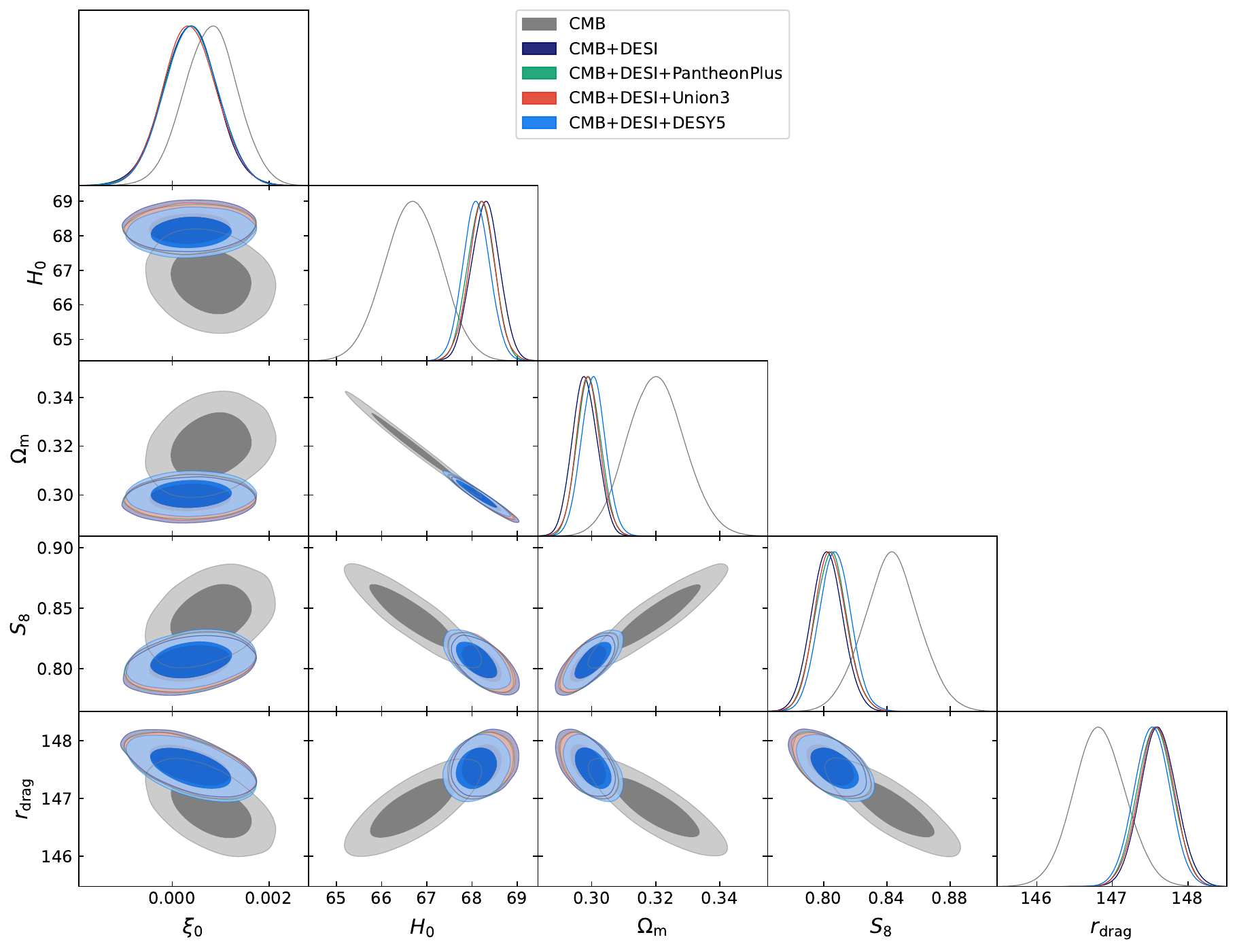}
	\caption{One-dimensional posterior distributions and two-dimensional joint contours for the most relevant parameters of the {\bf IVS2b} scenario, using several combinations of cosmological datasets.}
	\label{fig:IVS2b}
\end{figure*}		

\begin{figure*}
    \centering
    \includegraphics[width=0.45\textwidth]{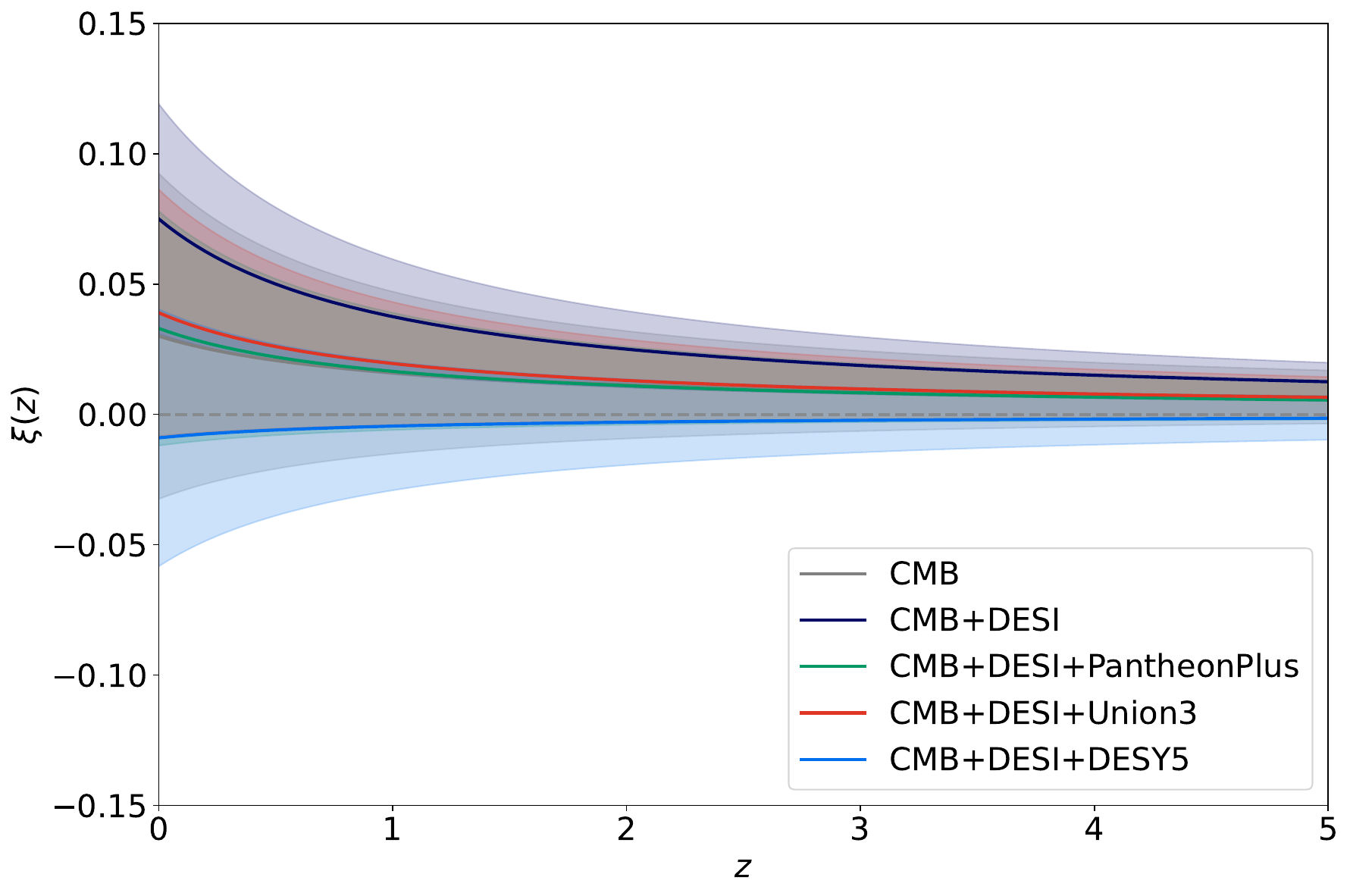}
    \includegraphics[width=0.45\textwidth]{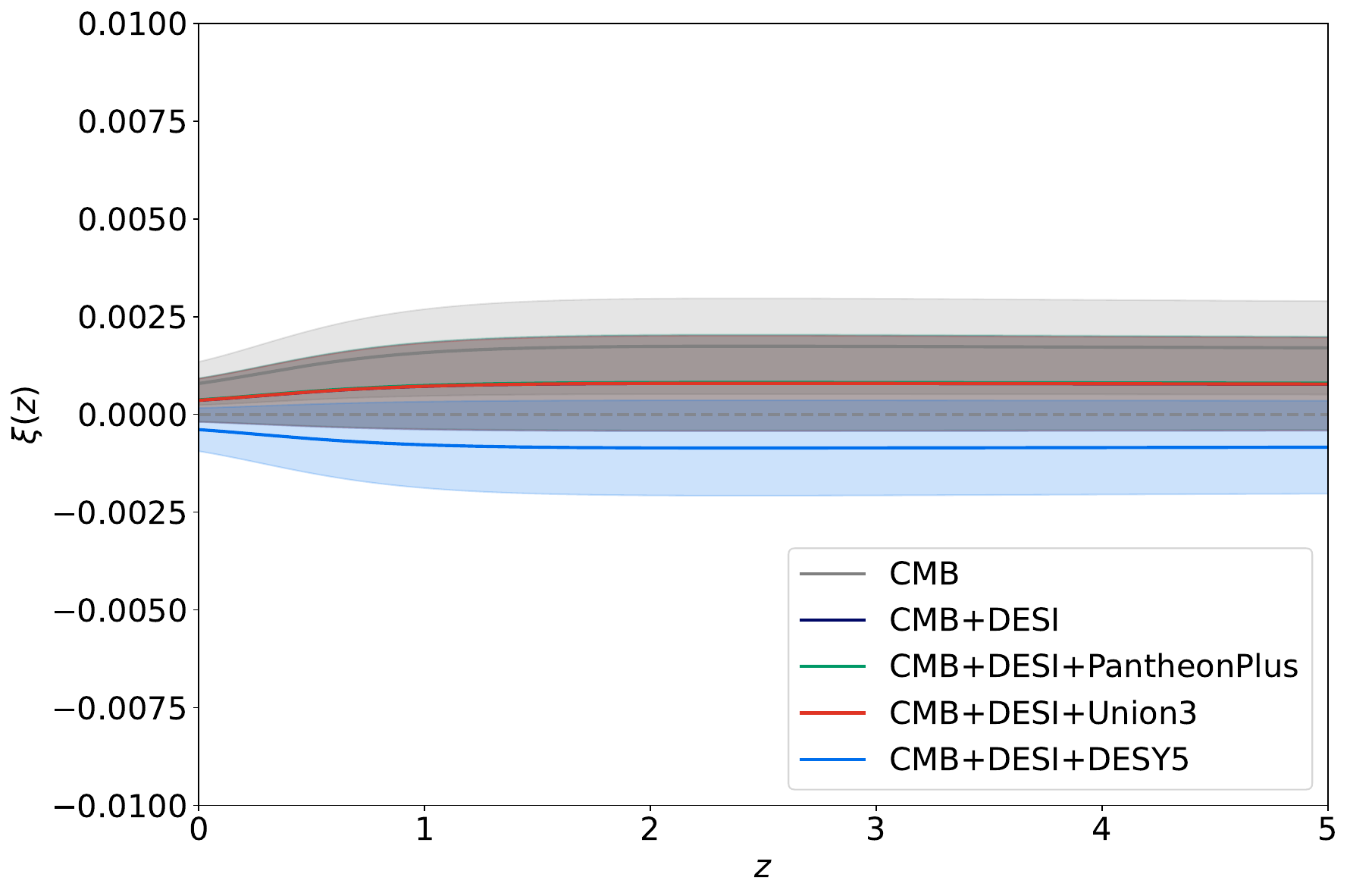}
    \caption{Evolution of $\xi(z)$ ($1 + z = a^{-1}$) with its 68\% CL region for {\bf IVS2a} (left plot) and {\bf IVS2b} (right plot), considering all the datasets. The horizontal dotted line in each plot corresponds to the no-interaction case (i.e., $\xi(z) = 0$).}
    \label{fig:xi(z)-ivs2a2b}
\end{figure*}

\subsection{IVS2a}

Table~\ref{table-IVS2a} and Fig.~\ref{fig:IVS2a} summarize the constraints on {\bf IVS2a}.

Considering the CMB-alone constraints, we find that although $\xi_0$ and $\xi_a$ are both non-zero at their mean values, their respective 68\% CL intervals include zero. Hence, no significant evidence for interaction is found in this case. The constraints on $H_0$ and $S_8$ are nearly identical to those of the non-interacting $\Lambda$CDM model, as indicated by Planck~\cite{Planck:2018vyg}.
In this scenario, the correlation structure among parameters plays a crucial role. As in $\Lambda$CDM, $\Omega_m$ and $H_0$ are strongly anti-correlated, which largely drives the degeneracies seen in the CMB-only contours. Additionally, the coupling parameter $\xi_0$ is positively correlated with $H_0$ and negatively with $\Omega_m$, so when $\xi_0$ is slightly greater than zero, $H_0$ tends to shift upward while $\Omega_m$ shifts downward. In contrast, $\xi_a$ exhibits the opposite behavior: it is positively correlated with $\Omega_m$ and negatively with $H_0$. Therefore, depending on the signs and magnitudes of $\xi_0$ and $\xi_a$, the addition of external datasets can produce compensating shifts in $\Omega_m$ and $H_0$, tightening the degeneracies and modifying the inferred cosmological values.

When DESI is combined with CMB, evidence of interaction begins to emerge. Specifically, $\xi_0 \neq 0$ at more than 68\% CL ($\xi_0 = 0.075^{+0.048}_{-0.039}$ at 68\% CL), and the dynamical nature of the coupling is supported by a non-zero value of $\xi_a$ at slightly more than 68\% CL ($\xi_a = -0.075^{+0.039}_{-0.048}$ at 68\% CL). Since $\xi_0 > 0$, the energy flows from CDM to DE, leading to a mild reduction in $\Omega_m$, which is compensated by a slightly higher value of $H_0$, yielding $H_0 = 68.95 \pm 0.48$ km/s/Mpc at 68\% CL.

When the three different SNIa datasets are individually added to CMB+DESI, we consistently find that both $\xi_0$ and $\xi_a$ remain consistent with zero within 68\% CL, although their mean values are non-zero. Therefore, these combinations do not provide significant evidence for interaction or for the dynamical nature of the coupling. The constraints on $H_0$ (and $S_8$) are slightly higher (lower) than those of the Planck $\Lambda$CDM baseline~\cite{Planck:2018vyg}.

However, according to the Bayesian evidence analysis, $\Lambda$CDM continues to be favored over this interacting scenario.

\subsection{IVS2b}

We now turn to the interacting scenario {\bf IVS2b}, in which the coupling parameter is dynamical and described by a single free parameter, as defined in Eq.~(\ref{xi(a)-2}). The constraints on this scenario are summarized in Table~\ref{table-IVS2b} and Fig.~\ref{fig:IVS2b}.

From CMB alone, we find mild evidence for interaction at more than 68\% CL, with $\xi_0 = 0.00079 \pm 0.00055$ at 68\% CL. However, this indication disappears once DESI data are added. The inclusion of SNIa datasets alongside CMB+DESI also yields results consistent with no interaction. In all cases, the mean values of $\xi_0$ remain slightly different from zero, but always close to it, and the 68\% CL intervals include the null value. Thus, apart from the CMB-only case, no statistically significant evidence for interaction is found.

In this scenario, the correlation structure is notably different from the other interacting models. The parameters $\Omega_m$ and $H_0$ remain strongly anti-correlated, as in the standard $\Lambda$CDM case. However, the coupling parameter $\xi_0$ shows no correlation with either $\Omega_m$ or $H_0$. As a result, the shifts observed in $\Omega_m$ and $H_0$ across different dataset combinations are not driven by the interaction itself, but rather by the intrinsic preferences of the datasets. This decoupling of $\xi_0$ from the key cosmological parameters distinguishes {\bf IVS2b} from the other scenarios discussed.

Regarding the Hubble constant, in the CMB-only case we observe a slight reduction in $H_0$ compared to the Planck estimate within the $\Lambda$CDM framework~\cite{Planck:2018vyg}. For the other combinations, $H_0$ increases slightly, with typical values around $H_0 \sim 68$ km/s/Mpc. For the $S_8$ parameter, we find that, except for the CMB-only case, the values are slightly reduced relative to Planck-$\Lambda$CDM, with $S_8 \sim 0.8$ in all combined analyses.

Finally, according to the Bayesian evidence analysis, $\Lambda$CDM remains favored over this interacting scenario.

In a similar fashion, we conclude this section with Fig.~\ref{fig:xi(z)-ivs2a2b}, which shows the time evolution of the coupling parameter $\xi(z)$ for both functional forms defined in Eqs.~(\ref{xi(a)-1}) and~(\ref{xi(a)-2}). Each plot also includes the 68\% CL region, derived from the full set of datasets considered in this analysis. 
From the left panel, corresponding to the two-parameter model, we observe a clear late-time evolution of $\xi(z)$ for most dataset combinations, except for CMB+DESI+DESY5, where the evolution is much less pronounced. In contrast, the right panel, corresponding to the one-parameter model, shows a much weaker evolution overall, indicating that $\xi(z)$ remains nearly constant over cosmic time across all dataset combinations.

\section{Summary and Conclusions}
\label{sec-summary}

In interacting scenarios between dark matter (DM) and dark energy (DE), the most crucial element is the interaction function, which governs the evolution of the dark components at both background and perturbative levels. The coupling function typically includes one or more coupling parameters, which are often assumed to be constant, i.e., not evolving with the expansion of the universe. While this assumption simplifies the analysis and avoids issues such as parameter degeneracies, there is no fundamental reason to exclude the possibility of a dynamical coupling. Allowing the coupling to vary with time can provide a richer phenomenology and potentially capture more subtle features of the dark sector.

In this article, we have studied interacting scenarios characterized by the interaction functions $Q = 3 \xi(a) H \rho_x$ and $Q = 3 \xi(a) H \frac{\rho_c \rho_x}{\rho_c + \rho_x}$, where the coupling function $\xi(a)$ is time-dependent and includes only one free parameter (although generalizations to multiple parameters are possible). Since the choice of the dynamical form of $\xi(a)$ is not unique, we considered two specific cases: {\bf (i)} $\xi(a) = \xi_0 + \xi_a (1 - a)$, which involves two free parameters, and {\bf (ii)} $\xi(a) = \xi_0 f(a)$, where $f(a) = 1 + \frac{1 - a}{a^2 + (1 - a)^2}$ is a divergence-free function, resulting in a single free parameter model. We constrained these interacting models using a combination of recent cosmological observations, including CMB, BAO from DESI, and three independent SNIa datasets: PantheonPlus, Union3, and DESY5. The results are presented in Tables~\ref{table-IVS1a}--\ref{table-IVS2b} and Figs.~\ref{fig:IVS1a}--\ref{fig:xi(z)-ivs2a2b}.

Our analysis reveals that for the {\bf IVS1a} scenario, which features a two-parameter coupling, there is robust evidence for an interaction at more than 95\% CL when CMB+DESI is combined with any of the SNIa datasets. In contrast, for {\bf IVS1b}, which uses the one-parameter divergence-free coupling, only mild evidence for an interaction (at just over 68\% CL) is found for specific dataset combinations. For {\bf IVS2a}, the results are less conclusive: a mild indication for an interaction (just above 68\% CL) results for CMB+DESI, but other combinations do not support such an interaction. In the case of {\bf IVS2b}, only the CMB-alone analysis yields a marginal preference for interaction, and even this remains below the 95\% CL threshold. These findings suggest that the observational outcome strongly depends on both the form of the interaction function and the parametrization chosen for $\xi(a)$.

It is important to emphasize that the goal of this work is not to confirm the existence of a dark sector interaction, but rather to explore its viability in light of the latest cosmological data. Our results highlight the sensitivity of the interaction constraints to both the model choice and the datasets used. Future high-precision surveys, especially those probing the low-redshift universe and the growth of structure, will be crucial for further testing these models and improving our understanding of the dark sector.

\acknowledgments
WY has been is supported by the National Natural Science Foundation of China under Grants No. 12175096, and Liaoning Revitalization Talents Program under Grant no. XLYC1907098. 
 OM acknowledges the financial support from the MCIU with funding from the European Union NextGenerationEU (PRTR-C17.I01) and Generalitat Valenciana (ASFAE/2022/020). OM is also supported by the Spanish MINISTERIO DE CIENCIA E INNOVACIÓN grants PID2023-148162NB-C22 and PID2020-113644GB-I00 and by the European ITN project HIDDeN (H2020-MSCA-ITN-2019/860881-HIDDeN) and SE project ASYMMETRY (HORIZON-MSCA-2021-SE-01/101086085-ASYMMETRY) and well as by the Generalitat Valenciana grant CIPROM/2022/69. OM acknowledges the financial support from the MCIU with funding from the European Union NextGenerationEU (PRTR-C17.I01) and Generalitat Valenciana (ASFAE/2022/020). SP acknowledges the partial support from  the Department of Science and Technology (DST), Govt. of India under the Scheme  ``Fund for Improvement of S\&T Infrastructure (FIST)'' [File No. SR/FST/MS-I/2019/41]. EDV is supported by a Royal Society Dorothy Hodgkin Research Fellowship. 
 This article is based upon work from the COST Action CA21136 - ``Addressing observational tensions in cosmology with systematics and fundamental physics (CosmoVerse)'', supported by COST - ``European Cooperation in Science and Technology''.
 
\bibliography{biblio}

\end{document}